\documentclass[aps,nofootinbib,preprintnumbers,twocolumn,superscriptaddress]{revtex4-1}

\usepackage{amssymb,amsmath,bm,natbib}
\usepackage{color}
\usepackage{slashed}
\usepackage{graphics}
\usepackage{graphicx}
\usepackage[utf8]{inputenc}
\usepackage[colorlinks = true,
            linkcolor = Maroon,
            urlcolor  = Maroon,
            citecolor = Maroon]{hyperref}
\usepackage{url}
\usepackage[dvipsnames]{xcolor}
\usepackage{dsfont}
\usepackage{float} 
\usepackage{cancel}
\usepackage{bm}
\usepackage[export]{adjustbox}
\usepackage{microtype}
\usepackage{soul}
\usepackage{multirow}
\usepackage[compat=1.1.0]{tikz-feynman}
\usepackage{caption}
\usepackage{subcaption}
\usepackage[version=4]{mhchem}

\newcommand{\MeV}{{\rm Me\kern -0.1em V}}
\newcommand{\GeV}{{\rm Ge\kern -0.1em V}}
\newcommand{\TeV}{{\rm Te\kern -0.1em V}}

\renewcommand{\i}{\ensuremath{\mathrm{i}}}

\usepackage{mfirstuc} % to uppercase the name
\newcommand{\addReviewer}[2]{
  \expandafter\newcommand\csname #1\endcsname[1]{{\textbf{ \color{#2} \capitalisewords{#1}:\,##1}}}
  \expandafter\newcommand\csname #1cor\endcsname[2]{{\color{#2} \capitalisewords{#1}:\,\st{##1}{\textbf{##2}}}}
  \expandafter\newcommand\csname #1color\endcsname{#2}
  \expandafter\newcommand\csname #1todo\endcsname[1]{{\todo[inline,color=white!70!#2, caption={}]{\textbf{\capitalisewords{#1}}: ##1}}}
}

\usepackage{soul,color}
\definecolor{chromeyellow}{rgb}{1.0, 0.65, 0.0}

\addReviewer{CAM}{blue}
\addReviewer{SP}{red}

\begin{document}
\preprint{ZU-TH 23/22, PSI-PR-22-15}

\title{A Flavour Inspired Model for Dark Matter}

\author{Claudio Andrea Manzari}
\email{claudioandrea.manzari@physik.uzh.ch}
\affiliation{Physik-Institut, Universit\"at Z\"urich, Winterthurerstrasse 190, CH-8057 Z\"urich, Switzerland}
\affiliation{Paul Scherrer Institut, CH-5232 Villigen PSI, Switzerland}

\author{Stefano Profumo}
\email{profumo@ucsc.edu}
\affiliation{Santa Cruz Institute for Particle Physics and Department of Physics, University of California, 1156 High St, Santa Cruz, CA 95060, United States of America}

%%%%%%%%%%%%%%%%%%%%%%%%%%%%%%%%

\begin{abstract}
The discrepancies between data on rare $b$-hadron decays, controlled by the underlying neutral-current transitions $b\to s\ell^+\ell^- (\ell = e, \mu)$, and the corresponding Standard Model predictions constitute one of the most intriguing hints for new physics. Leptoquarks are prime candidates to solve these anomalies and, in particular, the scalar leptoquark, $S_3$, triplet under $SU(2)_L$ with hypercharge $Y=-1/3$, provides a very good fit to data. Here, for the first time, we entertain the possibility that the same scalar leptoquark, responsible for the LFU anomalies, is the {\em portal} to a dark sector consisting of two additional vector-like fermions, one of which is a candidate for the cosmological dark matter. We study two scenarios, where the dark matter candidate belongs to an $SU(2)_L$ singlet and triplet respectively, and discuss the theory parameter space in the context of the dark matter candidate's relic density and prospects for direct and indirect dark matter searches. Direct detection rates are highly suppressed, and generically below the neutrino floor. Current observations with, and future prospects for, high-energy gamma-ray telescopes such as HESS and the Cherenkov Telescope Array are much more promising, as they already provide powerful constraints on the models under consideration, and will potentially probe the full parameter space in the future.
\end{abstract}
\maketitle

%%%%%%%%%%%%%%%%%%%%%%%%%%%%%%%

\section{Introduction}

Two of the most compelling and suggestive puzzles in modern particle physics are the particle nature of the cosmological dark matter and the hints for the violation of lepton flavour universality (LFU) in $B$ decays, as emerging from a variety of collider data in recent years (for recent reviews on the two topics see Refs.~\cite{Profumo:2017hqp,Profumo:2019ujg,CAMALICH20221,Crivellin:2021sff,Fischer:2021sqw}). While a continuing and extraordinarily intense program of searches for non-gravitational signals from dark matter has yet to deliver uncontroversial evidence for discovery of new physics \cite{Arcadi:2017kky}, mounting evidence for LFU violation makes it timely to investigate possible {\em connections} between the two sectors, with the hope of outlining and pinpointing expected signals especially for planned experiments and observatories. This, in short, is the spirit of the present note: the new particle responsible for the flavor anomalies (here a scalar leptoquark triplet) might be the mediator between the visible and the dark sector, the latter containing the particle making up the cosmological dark matter.

In the Standard Model (SM) of particle physics, LFU is satisfied by the gauge interactions and is only violated by the lepton Yukawa couplings. In recent years, several measurements have shown a coherent pattern of deviations from the SM predictions in rare b-hadron decays controlled by the underlying neutral current transition $b\to s\ell^+\ell^- (\ell = e, \mu)$~\cite{LHCb:2013ghj,LHCb:2014vgu,CMS:2014xfa,LHCb:2015svh,LHCb:2017avl,ATLAS:2018cur,CMS:2019bbr,LHCb:2020lmf,LHCb:2020gog}. Even though no single result exceeds the $5~\sigma$ threshold, conventionally considered a necessary condition to claim discovery of physics deviating from the SM, the combined analysis of the LFU ratios $R(K^{(*)})$, the branching ratio of $B_s\to\phi\mu^+\mu^-$ and angular observables in $B\to K^*\mu^+\mu^-$, show a preference for new physics (NP) scenarios over the SM hypothesis with an intriguing significance~\cite{Altmannshofer:2021qrr,Geng:2021nhg,Alguero:2021anc,Hurth:2021nsi,Isidori:2021vtc,Ciuchini:2021smi}. In particular, all these observables together point convincingly toward an effect in the di-muon channel.

Among the several SM extensions proposed in the literature to solve the $b\to s\ell^+\ell^-$ anomalies, leptoquarks are prime candidates~\cite{Alonso:2015sja, Calibbi:2015kma,Hiller:2016kry,Bhattacharya:2016mcc, Buttazzo:2017ixm,Barbieri:2015yvd,Barbieri:2016las, Calibbi:2017qbu,Crivellin:2017zlb, Crivellin:2017dsk,Bordone:2018nbg,Kumar:2018kmr, Crivellin:2018yvo,Crivellin:2019szf,Cornella:2019hct, Bordone:2019uzc,Bernigaud:2019bfy,Aebischer:2018acj,Fuentes-Martin:2019ign,Popov:2019tyc,Fajfer:2015ycq,Blanke:2018sro,deMedeirosVarzielas:2019lgb,Varzielas:2015iva,Crivellin:2019dwb,Saad:2020ihm,Saad:2020ucl,Gherardi:2020qhc,DaRold:2020bib,Heeck:2022znj}. Indeed, it turns out that the scalar leptoquark transforming as a triplet under $SU(2)_L$, which we will denote as $S_3$ hereafter, with couplings to muons, gives a very good fit to data. In the literature, scalar leptoquarks have been discussed as mediators to dark matter in Refs.~\cite{Choi:2018stw,Mandal:2018czf,Belanger:2021smw,Queiroz:2014pra,DEramo:2020sqv,Belfatto:2021ats,Baker:2021llj}. Here, for the first time, we entertain the possibility that the same scalar leptoquark $S_3$, responsible for the LFU anomalies, is the {\em portal} to a dark sector\footnote{We indicate the two vector-like fermions as belonging to a ``dark sector'' simply to emphasize that they are {\em additional} particles to the Standard Model, not because they are inert under the Standard Model gauge groups, since both actually do share quantum numbers with Standard Model particles.} consisting of two additional vector-like fermions\footnote{The Majorana or Dirac nature of these fermions is irrelevant for the phenomenology discussed here, so we do not specify it.}; of these, the lighter one, which we indicate with $\chi_L$ is neutral under $SU(3)_C$ and electromagnetism, and is the particle dark matter candidate; the second, colored one, $\chi_Q$, plays a key role in setting the thermal relic density and in shaping the outlook for perspective dark matter searches to test this scenario.

The remainder of this note is as follows: the next section describes the field content of the theory under consideration; the following section details the requirements to fit LFU violation observations on the model under consideration; sec.~\ref{sec:DM} elaborates on the dark sector and its leptoquark portal, and studies the implications of the model for the dark matter abundance and its direct and indirect search signals; finally, sec.~\ref{sec:concl} presents our discussion and conclusions.

\section{The Flavour Model}

The scalar leptoquark triplet, $S_3$, with quantum numbers $(3,3,-1/3)$ under the SM gauge group $SU(3)_C\times SU(2)_L\times U(1)_Y$, has only one Lorentz- and gauge-invariant coupling to SM fields:
\begin{align}
    \mathcal{L}_{\rm NP} = Y_{ij}\bar{Q}^c_i \i \tau_2 S_3^{\dagger} L_j + h.c.\,,
    \label{eq:LagB1}
\end{align}
where $Q$ and $L$ are the quark and lepton doublets under $SU(2)_L$ and our convention for $S_3$ is
\begin{align}
    S_3 = 
    \frac{1}{2}\begin{pmatrix}
    S^{-1/3} & \sqrt{2}S^{-4/3} \\
    \sqrt{2}S^{2/3} & -S^{-1/3}
    \end{pmatrix}\,.
\end{align}
In the following analysis we assume that the couplings $Y_{ij}$ are given in the down-quark basis. Therefore, after electroweak (EW) symmetry breaking the Lagrangian in Eq.~(\ref{eq:LagB1}) reads
\begin{align}
\begin{split}
    \mathcal{L}_{\rm NP} &= -\frac{Y_{ij}}{2\sqrt{2}}\bar{d}^c_i (S^{-4/3})^{\dagger} \ell_j -\frac{V_{ik}^*Y_{kj}}{4}\bar{u}^c_i (S^{-1/3})^{\dagger} \ell_j\\ &-\frac{Y_{ij}}{4}\bar{d}^c_i (S^{-1/3})^{\dagger} \nu_j + \frac{V_{ik}^*Y_{kj}}{2\sqrt{2}}\bar{u}^c_i (S^{2/3})^{\dagger} \nu_j + h.c.\,.
\end{split}
    \label{eq:LagB2}
\end{align}
where $V$ is the CKM matrix.\\
As stated in the introduction, $S_3$ is a well known candidate to solve the anomalies in rare $b$-hadron decays controlled by the underlying quark-level transition $b\to s\ell^+\ell^-$. Following  the conventions in Ref.~\cite{Capdevila:2017bsm}, we define the effective hamiltonian
\begin{align}
    \mathcal{H}_{\rm NP}^{b\to s\ell\ell} = -\frac{4G_F}{\sqrt{2}}V_{tb}V_{ts}^* \sum_i(C_iO_i+C_i'O_i')\,,
\end{align}
where the operators relevant for our discussion are
\begin{align}
\begin{split}
O_{9}^{ij} &= \frac{e^2}{16\pi^2}(\bar{s}\gamma_{\mu}P_{L}b)(\bar{\ell}_i\gamma^{\mu}\ell_j)\,,\\
O_{10}^{ij} &= \frac{e^2}{16\pi^2}(\bar{s}\gamma_{\mu}P_{L} b)(\bar{\ell}_i\gamma^{\mu}\gamma_5\ell_j)\,.
\end{split}
\end{align}
The analysis of data within the framework of effective theories has been performed by several theory group in the last years: we point the Reader to Refs.~\cite{Alguero:2019ptt,Altmannshofer:2021qrr,Ciuchini:2020gvn,Hurth:2021nsi} for the most complete and updated ones.

A scenario strongly favoured by data is $C_9^{22} = -C_{10}^{22} \simeq -0.39$, with new physics required only in the muon sector. Here and in the following we use the numerical results of Ref.~\cite{Altmannshofer:2021qrr} (note that apart from small numerical differences, this is a result common to all analyses).\\
The leptoquark triplet generates, at tree level, exactly the pattern of Wilson coefficients $C_9^{22} = -C_{10}^{22}$. Indeed, the couplings in Eq.~(\ref{eq:LagB2}) give the following contributions
\begin{align}
\begin{split}
C_9^{ij} = -C_{10}^{ij} = \frac{\sqrt{2}\,\pi}{32\, \alpha\, G_F V_{tb}V_{ts}^*}\frac{Y_{3i}Y_{2j}^*}{M_{S_3}^2}\,.
\end{split}
\end{align}
With the inputs of Ref.~\cite{Zyla:2020zbs} we find the following correlation between couplings and mass:
\begin{align}
Y_{3i}Y_{2j}^* =10^{-8} \left( \frac{M_{S_3}}{\rm GeV} \right)^2 \,.
\label{eq:FlavourCouplings}
\end{align}
%
%The requirement of perturbative couplings ($Y_{ij}\lesssim 4\pi$) gives an upper bound on the leptoquark mass $\sim 125\, \rm TeV$.\\
Note that current constraints from $B\to K\nu\nu$ are negligible~\cite{Belle:2017oht}, but future Belle II sensitivity will probe this scenario~\cite{Belle-II:2010dht}. \\
The most stringent bounds on leptoquarks (LQs) come from direct searches at colliders. Specifically, ATLAS and CMS search for LQs pair production with second and/or third fermion generations final state. The results of these searches, assuming that the pair production is dominated by QCD interactions ($gg\to LQ\,LQ^\dagger$), lead to model-independent bounds on the mass and branching fractions of the LQs~\cite{Angelescu:2018tyl}. The best lower limit on the mass of the scalar triplet is given by the 2-jet-2-muon final state and is in the range $1275-1530\; \rm GeV$ for a branching ratio between 0.5 and 1~\cite{CMS-PAS-LUM-18-002}.

A complementary constrain comes from the analysis of high-$p_T$ tails in Drell-Yan processes $pp\to\ell\ell$, which probe the $t$-channel exchange of a LQ. Recasting ATLAS searches for $Z^{\prime}$ at $36.1 fb^{-1}$, Ref.~\cite{Angelescu:2018tyl} shows the allowed region for the mass vs the couplings of $S_3$. In the whole mass range analysed, the maximum values allowed for its couplings to fermions are well above the ones required by Eq.~(\ref{eq:FlavourCouplings}).

In the following, we take a conservative stand and study dark matter scenarios for a LQ with a mass above $1.6\; \rm TeV$. In addition, we consider only the couplings to SM fermions required to realize the scenario $C_9^{22}=-C_{10}^{22}$ to explain the tension in $b\to s\ell^+\ell^-$ data: $Y_{32}$ and $Y_{22}$, which are taken to be real and equal, as the flavour data depend only on the combination $Y_{32}Y_{22}^*$. The couplings to first quark and lepton generations would anyway strongly affect the bounds from direct searches and are usually set to zero, while the remaining couplings to second and third generations would only change the branching ratios of the leptoquark.

\section{The Dark Matter model}\label{sec:DM}

We assume that the scalar leptoquark is the mediator between the "visible" SM sector and a dark sector containing the dark matter candidate. Specifically, we assume that the latter consists of two new vector-like fermions\footnote{The generalization to more than one family is straightforward.}, $\chi_Q$ and $\chi_L$. To prevent their mixing with SM fermions\footnote{This requirement avoids the mass bounds from Electroweak Precision Observables and direct searches at colliders.}, and the stability of the dark matter candidate, both $\chi_Q$ and $\chi_L$ are assumed to be odd under a $\mathcal{Z}_2$-type symmetry, whereas all other particles are chosen to be $\mathcal{Z}_2$-even. The interaction Lagrangian of our model reads:
\begin{align}
    \mathcal{L} = \mathcal{L}_{\rm SM} + Y_{ij}\bar{Q}^c_i \i \tau_2 S_3^{\dagger} L_j + \lambda\, \bar{\chi}^c_Q \i \tau_2 S_3^{\dagger}\chi_L + h.c.
\end{align}
We want the dark matter (DM) candidate to be color neutral, and we choose it to be a component of the $\chi_L$ multiplet. To suppress direct detection rates, we assume that $\chi_L$ transforms as singlet under $SU(3)_C$, and requiring gauge invariance, $\chi_Q$ as a triplet \cite{Profumo:2020zgi}. Furthermore, $\chi_L$ must have null hypercharge. In fact, DM with $Y\neq 0$ would have vector-like interactions with the $Z$ boson that produce spin-independent elastic cross sections orders of magnitude larger than current bounds~\cite{Cirelli:2005uq}. Gauge invariance again forces $\chi_Q$ to have the same hypercharge of the leptoquark, $Y_{\chi_Q} = -1/3$. Finally, we require the DM candidate to be electromagnetically neutral. This, together with the requirement of null hypercharge, forces $\chi_L$ to be in a odd representation of $SU(2)_L$: a singlet, a triplet, a quintuplet, and so on. In the following, we discuss the two scenarios with the lowest dimensional representations for $\chi_L$ and $\chi_Q$, as shown in Table~\ref{Tab:DMqn}. Note, that higher dimensional representations are also possible. In these cases, we expect the results discussed in the following to be shifted to higher DM masses, due to the smaller mass splitting between the charge and neutral components of the multiplets, up to the point where perturbativity of the $SU(2)_L$ coupling is lost at energies below the Planck scale for large-dimensional representations~\cite{Cirelli:2005uq}.  

\begin{center}
\begin{table}
\begin{tabular}{|c|c|c c c|} 
 \hline
 & & $SU(2)_L$ & $U(1)_Y$ & $SU(3)_C$ \\
 \hline\hline
 \multirow{2}{7em}{{\color{Maroon}{SINGLET}}} & $\chi_L$ & 1 & 0 & 1\\ 
 & $\chi_Q$ & 3 & -1/3 & 3\\
 \hline
 \multirow{2}{7em}{{\color{Maroon}{TRIPLET}}} & $\chi_L$ & 3 & 0 & 1\\ 
 & $\chi_Q$ & 1 & -1/3 & 3\\
 \hline
\end{tabular}
\caption{Possible quantum numbers for the Dark Sector\label{Tab:DMqn}}
\end{table}
\end{center}

\subsection{SINGLET}

\begin{figure*}[h]
	\centering
	\begin{subfigure}[b]{0.3\textwidth}
		\includegraphics[width=1.0\textwidth]{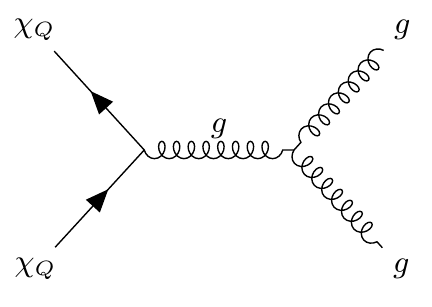}
		\caption{}
		\label{fig:QQgg}
	\end{subfigure}
	\;
	\begin{subfigure}[b]{0.3\textwidth}
		\includegraphics[width=0.75\textwidth]{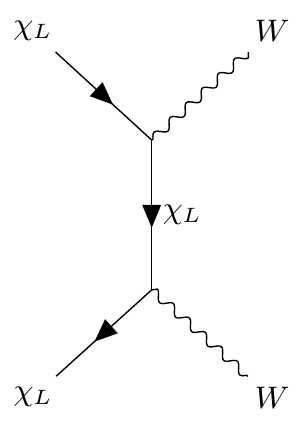}
		\caption{}
		\label{fig:LLWW}
	\end{subfigure}
	\\
	\begin{subfigure}[b]{0.3\textwidth}
		\includegraphics[width=0.75\textwidth]{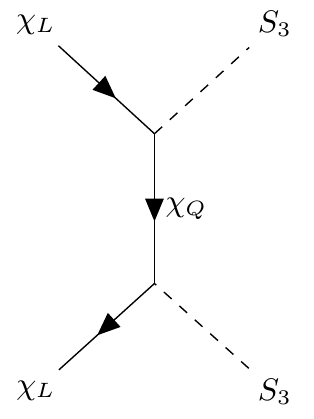}
		\caption{}
		\label{fig:LLSS}
	\end{subfigure}
	\;
	\begin{subfigure}[b]{0.3\textwidth}
		\includegraphics[width=1.0\textwidth]{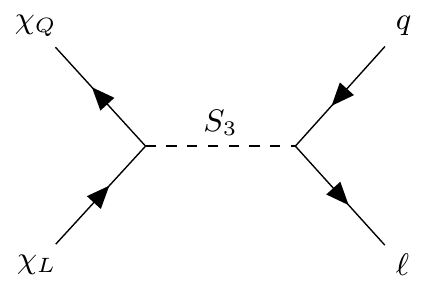}
		\caption{}
		\label{fig:LQql}
	\end{subfigure}
	\;
	\begin{subfigure}[b]{0.3\textwidth}
		 \includegraphics[width=0.75\textwidth]{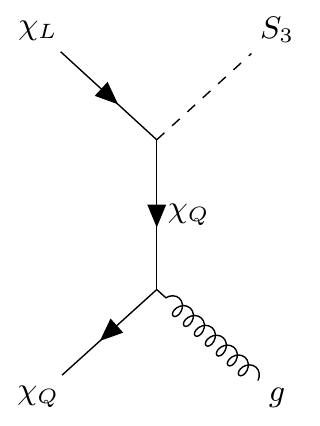}
		\caption{}
		\label{fig:LQSg}
	\end{subfigure}
	\caption{Feynman Diagrams of processes relevant for the determination of the Dark Matter relic density. Diagrams (a) and (b) show the self-annihilation of $\chi_L$ and $\chi_Q$ into SM gauge bosons, diagram (c) $\chi_L$ self-annihilation into two leptoquarks and diagrams (d) and (e) the co-annihilation of $\chi_L$ and $\chi_Q$.}
	\label{fig:FD}
\end{figure*}

We start with the case where $\chi_L$ is an $SU(2)_L$ singlet with hypercharge 0. With these quantum numbers, $\chi_L$ does not couple to any SM field. The case with $\lambda=0$ is trivial, as the DM would be stable and completely inert, with a relic density solely dictated by initial conditions. %In this situation, the correct relic density can always be achieved by a suitable choice of initial conditions. 
Turning on the coupling $\lambda$, the relic density is dominantly determined by the annihilation processes shown in Fig.~\ref{fig:LLSS}, the co-annihilation diagrams in Fig.~\ref{fig:LQql} and~\ref{fig:LQSg}, and the annihilation processes for the $\chi_Q$ shown in Fig.~\ref{fig:QQgg}: if $\chi_Q$ is close-enough in mass such that at the freeze-out of the $\chi_L$ its abundance is non-negligible, the relic density will result from a large set of inter-connected processes involving $\chi_L\chi_L\longleftrightarrow SM$, $\chi_L\chi_Q\longleftrightarrow SM$ and $\chi_Q\chi_Q\longleftrightarrow SM$ processes, where $\chi_Q$ indicates any member of the associated $SU(2)_L$ triplet and $SM$ any possible combination of Standard Model particles.

The scattering rates associated with the processes shown in fig.~\ref{fig:FD} are sensitive to the coupling $\lambda$, the leptoquark mass, $M_{S_3}$ and the masses of vector-like fermions. In the present study, we have implemented the model's Lagrangian and resulting Feynman rules in the {\tt micrOMEGAs} package \cite{Barducci:2016pcb}, and numerically computed all of the relevant annihilation and co-annihilation rates. We show our results in Fig.~\ref{fig:Relic}, top, which shows the relic density of dark matter as a function of $M_{\chi_L}$, for a fixed coupling, $\lambda=2$, and different values of $M_{S_3}$ and of the mass splitting between the $\chi_Q$  and the $\chi_L$, which we indicate with $\Delta$.

For zero mass splitting, $\Delta = M_{\chi_L} - M_{\chi_Q} = 0$, the coannihilation processes and the $\chi_Q\chi_Q$ annihilation processes are very efficient, driving to low values the thermal relic density of the $\chi_L$ as well. Furthermore, the lighter the leptoquark, the larger the co-annihilation processes, whose rates scale as $\propto \lambda^2/M_{S_3}^4$, and the smaller the thermal relic density: this is why the blue lines are much lower than the yellow lines in the plot. An increment to the mass splitting causes the suppression of the co-annihilation processes, as the thermal relic density of the $\chi_Q$ at $\chi_L$ freeze-out, $T_{\rm f.0.}\approx M_{\chi_L}/20$, is Boltzmann-suppressed by an approximate factor
\begin{align*}
    \propto \frac{\exp \left(-M_{\chi_Q}/T_{\rm f.o.}\right)}{\exp\left(-M_{\chi_L}/T_{\rm f.o.}\right)}\approx\exp \left(-20\Delta/M_{\chi_L}\right).
\end{align*}
In the figure, we find in fact that for light enough $\chi_L$, such that $20\Delta/M_{\chi_L}\sim 1$, the relic density steeply increases, as the coannihilation network with $\chi_Q$ is shut off.

There are two additional interesting features in the behavior of the thermal relic density as a function of $M_{\chi_L}$: the dip at $M_{\chi_L}+M_{\chi_Q}\sim M_{S_3}$ corresponds to the co-annihilation diagram in Fig.~\ref{fig:LQql} being resonantly enhanced because of the quasi-on-shell exchange of the scalar leptoquark; similarly, the threshold at $M_{\chi_L}+M_{\chi_Q}\sim M_{S_3}$ signals that the diagram in Fig.~\ref{fig:LQSg} is kinematically allowed, also driving the thermal relic density to lower values. At the lowest end of the mass range we consider, we note that we stop the lines where collider searches would have unveiled strongly interacting particles associated with the $\chi_Q$ triplet.

In sum, the observed relic density, $\Omega h^2=0.12\pm 0.0012$~\cite{Planck:2018vyg}, can be obtained only with DM masses above 2 TeV for small mass splittings between $\chi_Q$ and $\chi_L$. Increasing the mass splitting, solutions with lower masses become possible, including around 1 TeV, and around half the mass of the scalar leptoquark. A higher(lower) value for $\lambda$ would decrease(increase) the relic density, without changing the general behaviour shown in Fig.~\ref{fig:Relic}. In general, the relic density {\em increases} as the masses of vector-like fermions become larger, with the two  exceptions alluded to above:
\begin{itemize}
    \item For $M_{\chi_L} + M_{\chi_Q} = M_{S_3}$ the relic density drops significantly because of the enhancement of the process shown in Fig.~\ref{fig:LQql} where the leptoquark can be on-shell;
    
    \item For $M_{\chi_L} = M_{S_3}$ the relic density drops since the final state in Fig.~\ref{fig:LLSS} and Fig.~\ref{fig:LQSg} can be produced on-shell, enhancing the contribution of these processes.
\end{itemize}

Moving on to dark matter direct searches, in the top panel of Fig.~\ref{fig:DirectBounds}, we show the prediction for the spin-independent dark matter-proton cross section, as a function of $M_{\chi_L}$ and for different value of $M_{S_3}$ and $\Delta$ (and thus $M_{\chi_Q}$). For each point, the value of the coupling $\lambda$ has been fixed in order to reproduce the correct relic density. We find that the predicted cross section is too small to be probed by current experiments~\cite{Zyla:2020zbs} and is always below the neutrino coherent scattering floor~\cite{Strigari:2009bq}. This is a direct consequence of the requirement $Y_{\chi_L} = 0$ which avoids the vectorial coupling to the $Z$ boson, leaving, to leading order, the loop-suppressed diagram with $\chi_Q-S_3$ in the loop; as visible in the orange lines, we find that for $M_{\chi_L}\simeq 2.5$ TeV cancellations occur and the scattering cross section goes to zero in a certain interval of dark matter particle masses.

As far as indirect searches are concerned, for the same combinations of $M_{S_3}$ and $\Delta$, we show the predicted pair-annihilation rate as a function of $M_{\chi_L}$ in the top panel of Fig~\ref{fig:IndirectBounds}. Here, the situation is far more promising. The relevant process is the pair production of an on-shell or off-shell leptoquark pair in the final state, which subsequently decays into SM quarks and leptons. With the minimal couplings needed to solve the flavour anomalies, this leads to $\mu^-,\,\nu_{\mu},\,b,\,t,\,c,\,s$ and respective anti-particles in the final state. We find that by far the largest contributions stem from two leptoquarks on-shell, which produces a pair-annihilation cross section comparable to current experimental limits. Here we show both current constraints from 6 year-long observations of nearby dwarf spheroidal galaxies (dSph) from the Fermi-LAT, for a $\bar b b$ final state \cite{Fermi-LAT:2015att}, closest to the spectrum expected from leptoquark decay, and from observations of the Galactic center from HESS \cite{HESS:2016mib}, for the same final state; additionally, we show projections for Fermi LAT (again for the $\bar b b$ final state) that build on anticipated additional dSph discoveries with LSST \cite{LSSTDarkMatterGroup:2019mwo}, and projections for the Cherenkov Telescope Array (CTA) for both the $\bar b b$ and the $WW$ final state (we refer the Reader to Ref.~\cite{CTA:2020qlo} for details and assumptions for the CTA analysis; we use here their predicted sensitivity as shown in fig.~14). 

For light leptoquark masses, a large portion of $M_{\chi_L}$ can be probed by HESS~\cite{HESS:2021pgk}, while future bounds from the Cherenkov Telescope Array (CTA) on $b\bar{b}$ final states~\cite{CTA:2020qlo} will be very powerful in probing the whole mass spectrum of our model, as long as leptoquarks can be produced on shell. On the other hand, we find that the current bounds and the future projections from  Fermi LAT cannot probe, nor will they probe in the future, this model.  

\subsection{TRIPLET}

A second viable scenario is one where $\chi_L$ is an $SU(2)_L$ triplet with hypercharge 0: the triplet $\chi_L = (\chi_L^+, \chi_L^0, \chi_L^-)$. There is in this case an additional class of processes relevant for the relic density involving the coupling to $W$ bosons, shown in Fig.~\ref{fig:LLWW}; the resulting contribution depends only on the mass of the fermions in the dark sector, as the couplings to gauge bosons are fixed by the choice of the representation. As before, strongly interacting processes (Fig.~\ref{fig:QQgg}) drive the pair-annihilation rate for $\chi_Q$, while a second class of processes involves the coupling with a leptoquark,  shown in Figs.~\ref{fig:LLSS},~\ref{fig:LQql} and~\ref{fig:LQSg}. The latter processes are therefore sensitive also to the coupling $\lambda$ and to the leptoquark mass, $M_{S_3}$. 

\begin{figure*}
    \centering
     \includegraphics[width=0.6\textwidth]{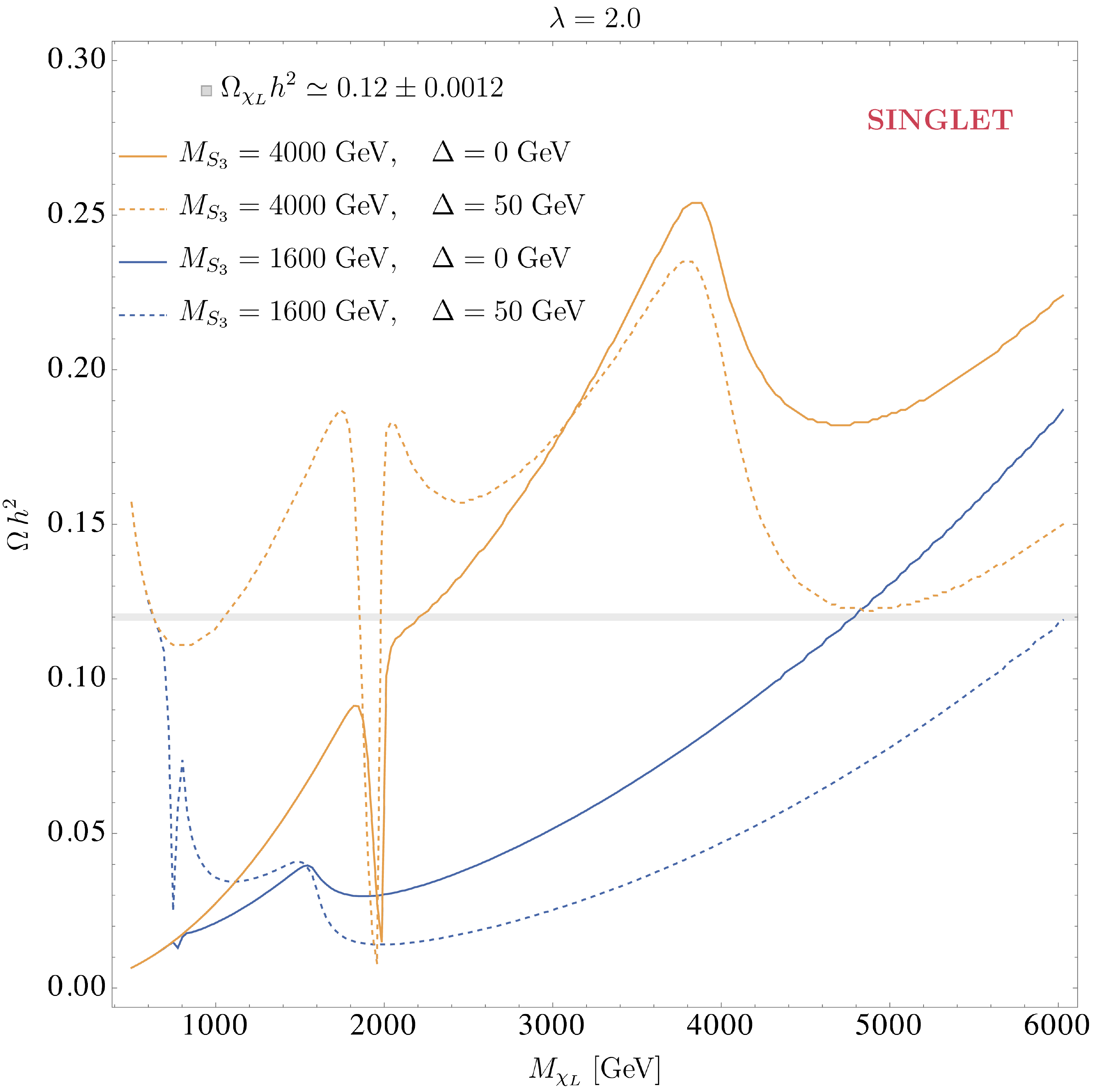}\\
   \includegraphics[width=0.6\textwidth]{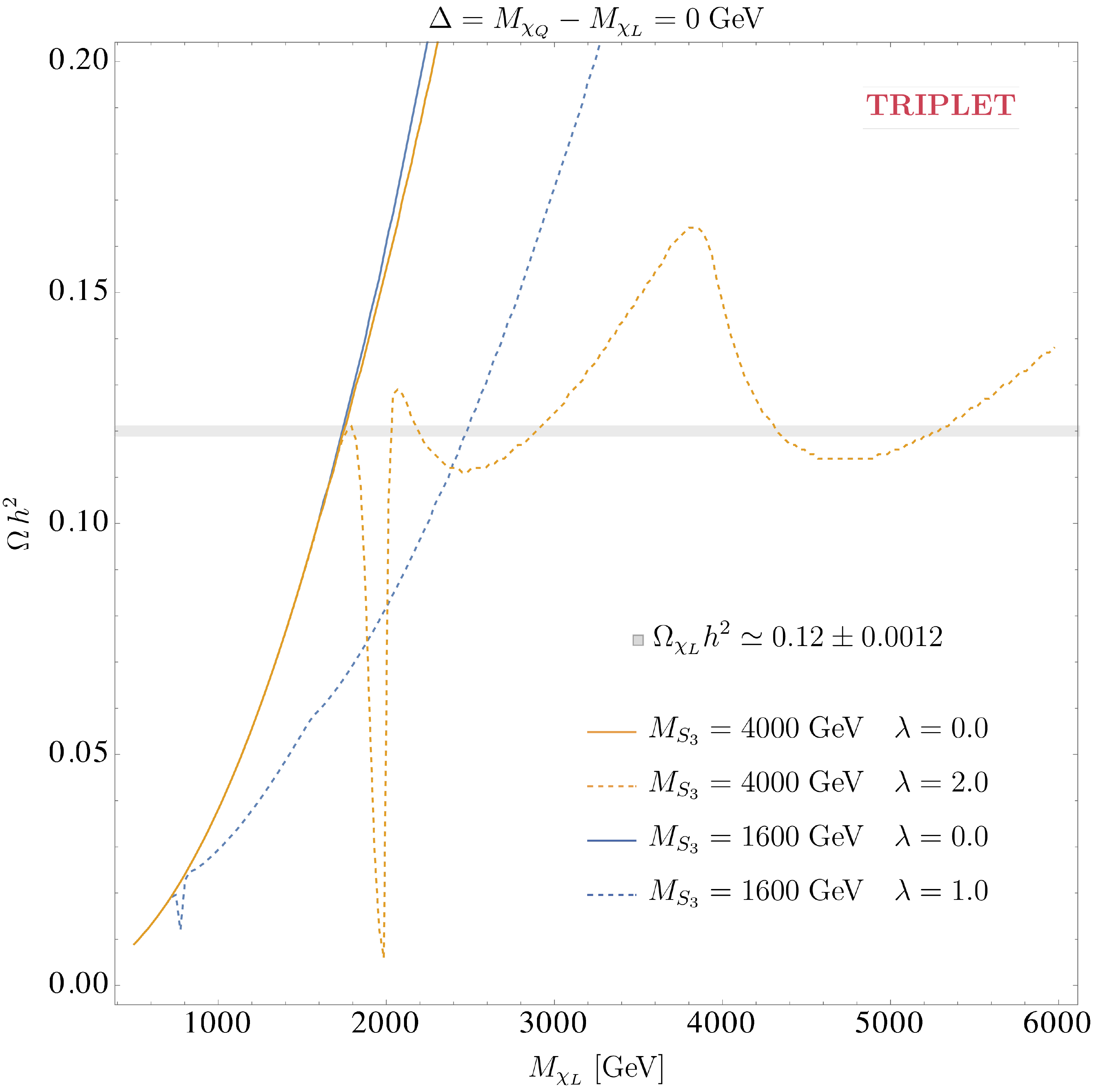}
    \caption{Relic abundance as a function of the dark matter mass for different values of the leptoquark mass $M_{S_3}$ and coupling. $\lambda$. The current experimental average for $\Omega\, h^2$ is also shown as a grey band.}
    \label{fig:Relic}
\end{figure*}

\begin{figure*}
    \centering
    \includegraphics[width=0.8\textwidth]{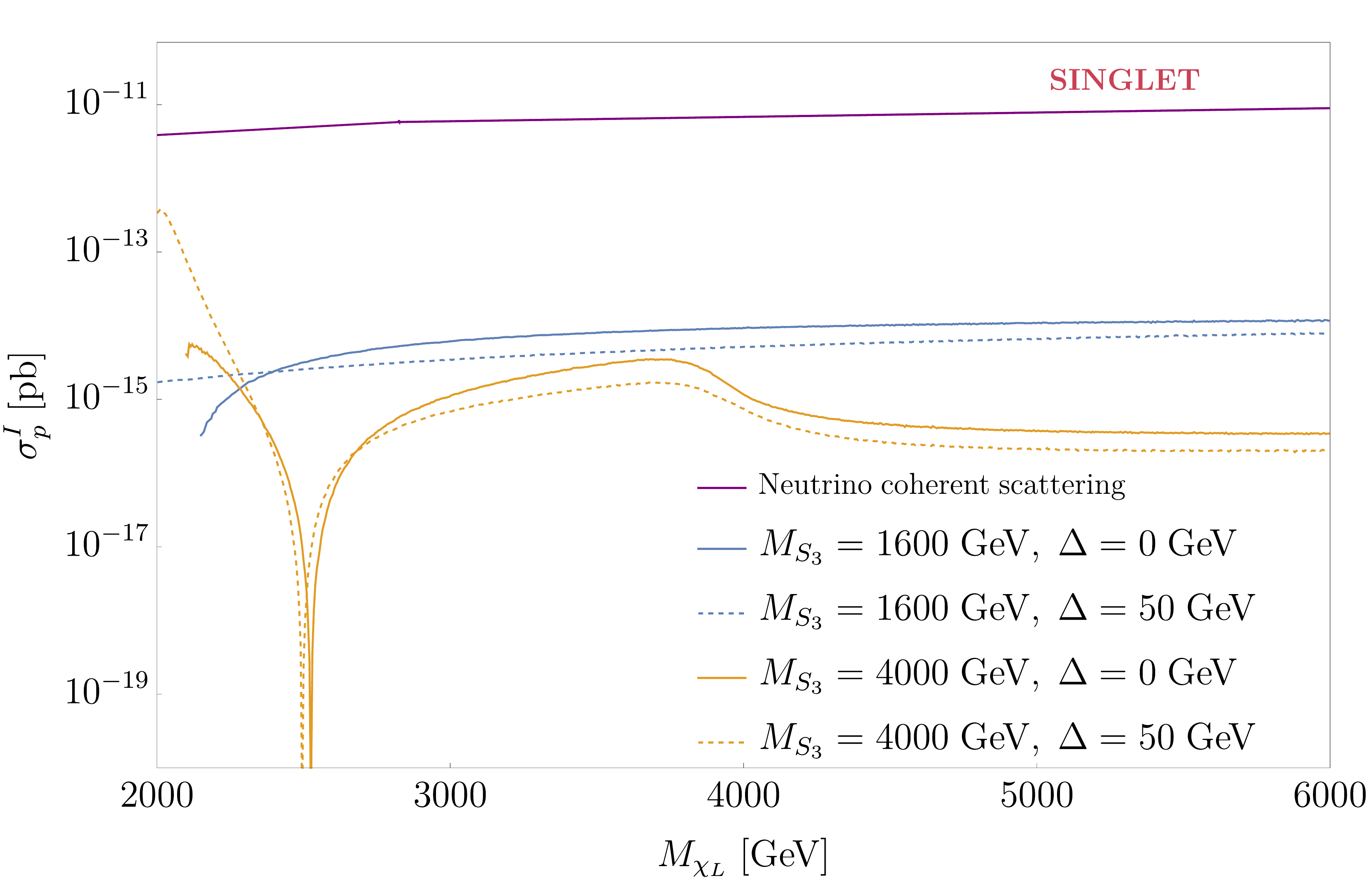}\\
        \includegraphics[width=0.8\textwidth]{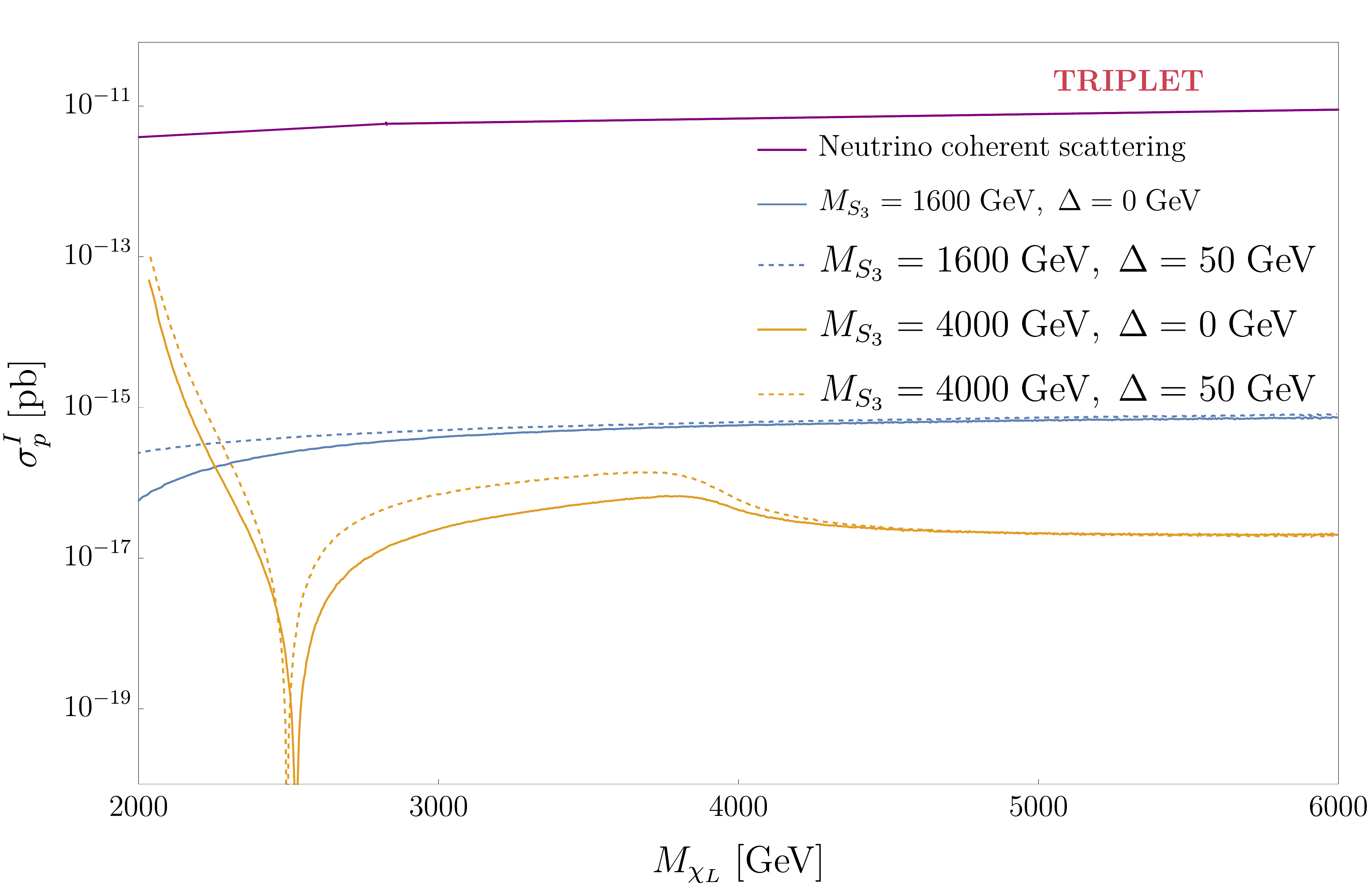}
    \caption{Spin-independent dark matter-proton cross section as a function of the dark matter mass, predicted requiring the observed relic abundance for different values of the leptoquark mass and mass splitting.}
    \label{fig:DirectBounds}
\end{figure*}

Fig.~\ref{fig:Relic}, bottom, shows the relic density of dark matter as a function of the triplet mass for different values of $M_{S_3}$ and $\lambda$, assuming $\Delta = M_{\chi_L} - M_{\chi_Q} =0$, thus complete degeneracy and hence maximal effect of coannihilation processes between $\chi_Q$ and $\chi_L$, for non-vanishing $\lambda$. The observed value of the cosmological dark matter is also shown as a grey band. As expected, for $\lambda=0$ the relic density is almost independent on the mass of the leptoquark (the observable small dependence is due to the fact that $S_3$ can still contribute at loop level via its couplings with SM fields). For $\lambda=0$ coannihilation are shut off, and the relic density is almost exclusively set by weak processes within the $\chi_L$ triplet, schematically shown in Fig.~\ref{fig:LLWW}. Processes $\chi_L^{0(+)} \chi_L^{0(-)} \to W^+ W^-$ with the exchange of a $\chi_L^{\pm(0)}$ in the $t$-channel are also very efficient. Since these contributions cannot be turned off, we can set a lower limit on the mass of the triplet by requiring the correct relic density. For $\Delta=0$ we find $M_{\chi_L}\gtrsim 1750\, \rm GeV$, while for a decoupling color singlet we find $M_{\chi_L}\gtrsim 1600\, \rm GeV$.

For $\lambda \neq 0$, the dependence of $\Omega h^2$ on the triplet mass becomes more complicate, as can be seen in Fig.~\ref{fig:Relic}, and a discussion similar to the singlet case applies. It is easy to generalize the results shown in Fig.~\ref{fig:Relic} to different values of $\Delta$ and $\lambda$: the effect of a larger $\Delta$ is to shift the dashed lines to the left, while the effect of a larger(smaller) $\lambda$ is to shift the same lines upwards(downwards).

\begin{figure*}
    \centering
    \includegraphics[width=0.8\textwidth]{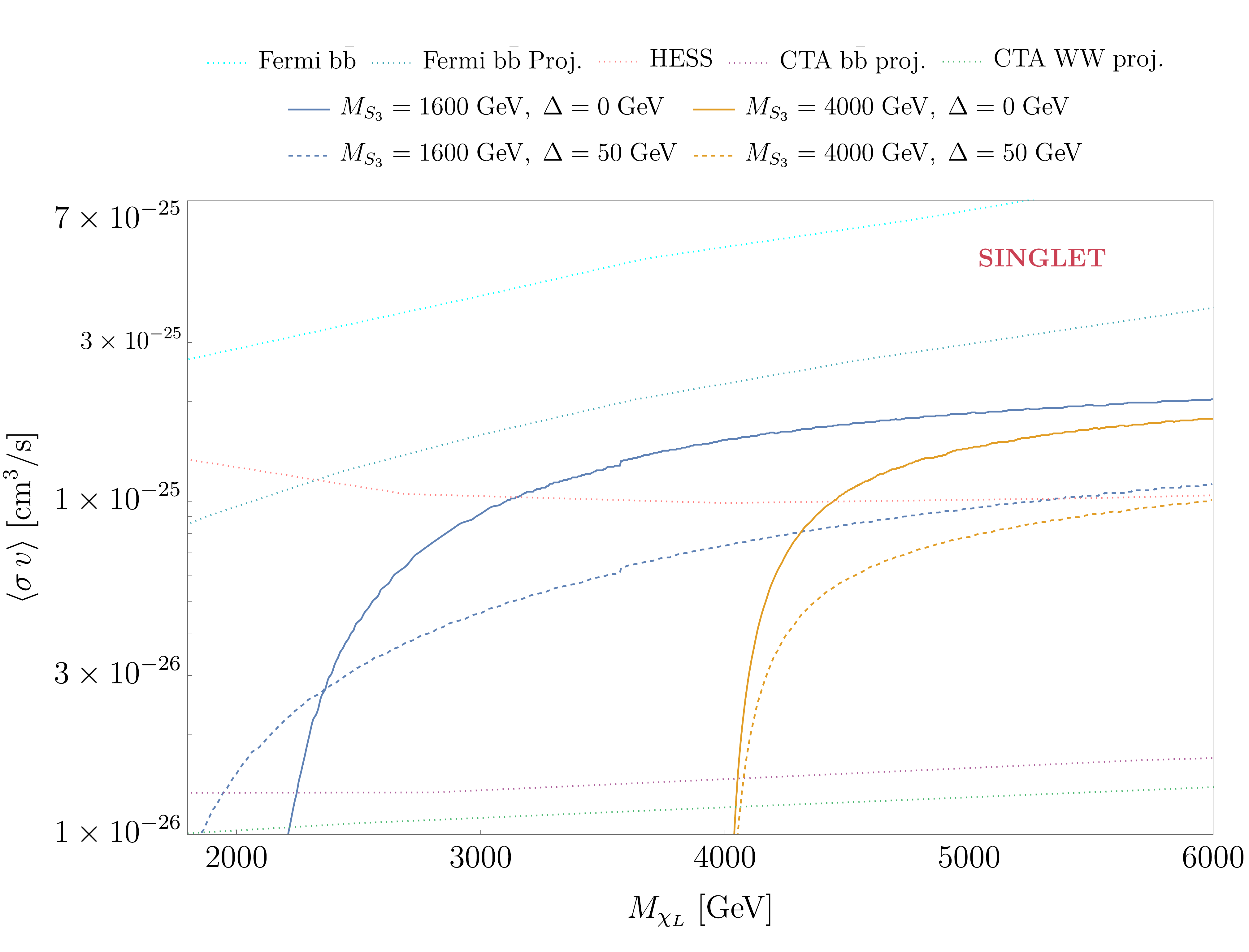}\\
    \includegraphics[width=0.8\textwidth]{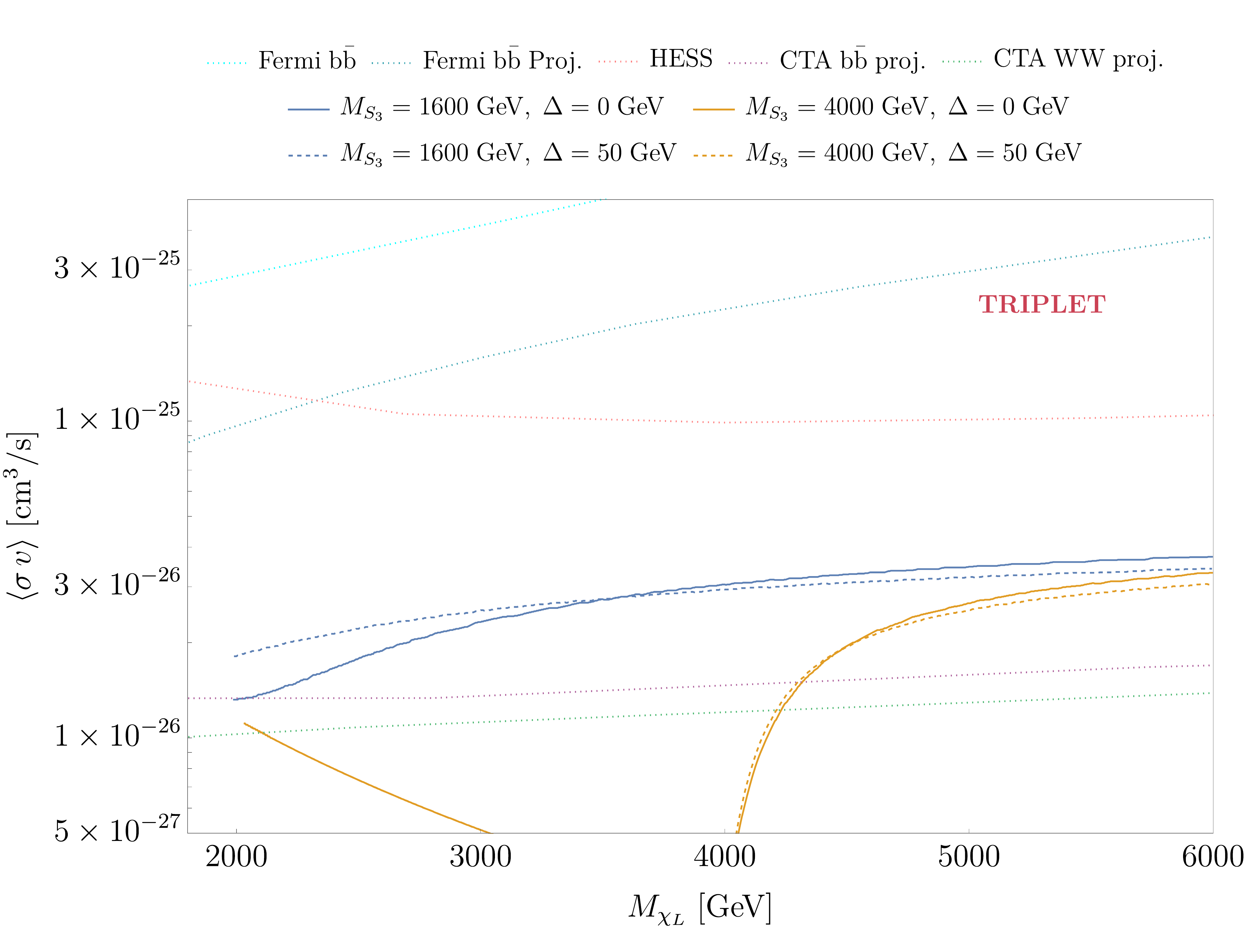}
    \caption{Pair-annihilation cross section as a function of the dark matter mass, predicted requiring the observed relic abundance for different values of the leptoquark mass an mass splitting.}
    \label{fig:IndirectBounds}
\end{figure*}

Fixing the value of the coupling $\lambda$ to reproduce the correct relic density, we show in Fig.~\ref{fig:DirectBounds} the prediction for the spin-independent dark matter-proton cross section, $\sigma^I_p$, and in in Fig~\ref{fig:IndirectBounds} the predicted pair-annihilation rate $\langle \sigma v\rangle$, as a function of the triplet mass and for different value of $M_{S_3}$ and $\Delta$. As for the singlet, we find that current experimental limits on $\sigma^I_p$~\cite{Zyla:2020zbs} cannot probe the predicted cross section which lies below the neutrino coherent scattering floor~\cite{Strigari:2009bq}. On the other hand, we find that for light leptoquark masses, the anticipated sensitivity of the Cherenkov Telescope Array on $b\bar{b}$ and $W^+W^-$~\cite{CTA:2020qlo} will potentially test a large range of the triplet mass. Here, the most relevant processes are the pair production of $W^+W^-$ and $S_3S_3$, for a light and a heavy triplet, respectively. In general, CTA bounds become relevant when the two leptoquarks in the final state can be produced on-shell. The contribution from processes with off-shell leptoquarks are not included in Fig.~\ref{fig:IndirectBounds}, however we checked that they are suppressed by several orders of magnitude with respect to the on-shell ones that are part of the results we show. 

\section{Discussion and Conclusion}\label{sec:concl}
Based on persistent experimental signals of lepton flavor universality violation, and on the as-persistent lack of any evidence in favor of any particular particle dark matter candidate, in this study we entertain the following scenario: we posit that the dark matter belongs to a {\em dark sector} whose portal to the {\em visible}, Standard Model sector is a scalar leptoquark $SU(2)_L$ triplet that also provides a solution and an explanation to the flavor anomalies.

We showed that the dark sector must include new fields charged under $SU(2)_L$ and $SU(3)_C$; we assumed the new fields to consist of two vector-like multiplets, one singlet under $SU(3)_C$, denoted as $\chi_L$, that contains the dark matter candidate of the theory, and one triplet under $SU(3)_C$, denoted as $\chi_Q$. 
We considered the two lowest-dimensional $SU(2)_L$ representation assignments for $\chi_L$ such that the resulting hypercharge is 0, as required by direct dark matter searches.

We computed in detail the thermal relic density of the $\chi_L$, and studied which processes drive it, and where and how such relic density corresponds to the observed abundance of the cosmological dark matter.

Finally, we explored direct and indirect dark matter searches: broadly, direct detection rates are loop-suppressed and consistently below the neutrino floor, thus unaccessible with current detector technology. For indirect detection, the model is not accessible with Fermi LAT observations, but it is at present constrained by higher-energy ground-based telescopes such as HESS, and will be probed in most of the relevant parameter space by the next generation Cherenkov Telescope Array.

If indeed lepton flavor universality violation is a signal of new physics, and such new physics sector includes a dark matter candidate, we show here, under certain unavoidable assumptions, what the resulting phenomenology might unfold. Our main conclusion is that the expectation is for a heavy, TeV or heavier, dark matter candidate with a very weak coupling to nucleons, but with large-enough pair-annihilation rates so that indirect detection with very high-energy gamma rays is the most promising path to discovery.

\section*{Acknowledgements}
We gratefully acknowledge W. Altmannshofer for the useful comments and suggestions on this draft. C.A.M. warmly thanks W. Altmannshofer and UC Santa Cruz for the hospitality. The work of C.A.M. is supported by the Swiss National Science Foundation (PP00P2\_176884).
S.P. is partly supported by the U.S.\ Department of Energy grant number de-sc0010107.  

\newpage
\makeatletter\onecolumngrid@push\makeatother
~
\vspace{1.5in}

\clearpage
\makeatletter\onecolumngrid@pop\makeatother
\bibliography{bibliography}

\begin{thebibliography}{76}%
\makeatletter
\providecommand \@ifxundefined [1]{%
 \@ifx{#1\undefined}
}%
\providecommand \@ifnum [1]{%
 \ifnum #1\expandafter \@firstoftwo
 \else \expandafter \@secondoftwo
 \fi
}%
\providecommand \@ifx [1]{%
 \ifx #1\expandafter \@firstoftwo
 \else \expandafter \@secondoftwo
 \fi
}%
\providecommand \natexlab [1]{#1}%
\providecommand \enquote  [1]{``#1''}%
\providecommand \bibnamefont  [1]{#1}%
\providecommand \bibfnamefont [1]{#1}%
\providecommand \citenamefont [1]{#1}%
\providecommand \href@noop [0]{\@secondoftwo}%
\providecommand \href [0]{\begingroup \@sanitize@url \@href}%
\providecommand \@href[1]{\@@startlink{#1}\@@href}%
\providecommand \@@href[1]{\endgroup#1\@@endlink}%
\providecommand \@sanitize@url [0]{\catcode `\\12\catcode `\$12\catcode
  `\&12\catcode `\#12\catcode `\^12\catcode `\_12\catcode `\%12\relax}%
\providecommand \@@startlink[1]{}%
\providecommand \@@endlink[0]{}%
\providecommand \url  [0]{\begingroup\@sanitize@url \@url }%
\providecommand \@url [1]{\endgroup\@href {#1}{\urlprefix }}%
\providecommand \urlprefix  [0]{URL }%
\providecommand \Eprint [0]{\href }%
\providecommand \doibase [0]{http://dx.doi.org/}%
\providecommand \selectlanguage [0]{\@gobble}%
\providecommand \bibinfo  [0]{\@secondoftwo}%
\providecommand \bibfield  [0]{\@secondoftwo}%
\providecommand \translation [1]{[#1]}%
\providecommand \BibitemOpen [0]{}%
\providecommand \bibitemStop [0]{}%
\providecommand \bibitemNoStop [0]{.\EOS\space}%
\providecommand \EOS [0]{\spacefactor3000\relax}%
\providecommand \BibitemShut  [1]{\csname bibitem#1\endcsname}%
\let\auto@bib@innerbib\@empty
%</preamble>
\bibitem [{\citenamefont {Profumo}(2017)}]{Profumo:2017hqp}%
  \BibitemOpen
  \bibfield  {author} {\bibinfo {author} {\bibfnamefont {S.}~\bibnamefont
  {Profumo}},\ }\href {\doibase 10.1142/q0001} {\emph {\bibinfo {title} {{An
  Introduction to Particle Dark Matter}}}}\ (\bibinfo  {publisher} {World
  Scientific},\ \bibinfo {year} {2017})\BibitemShut {NoStop}%
\bibitem [{\citenamefont {Profumo}\ \emph {et~al.}(2019)\citenamefont
  {Profumo}, \citenamefont {Giani},\ and\ \citenamefont
  {Piattella}}]{Profumo:2019ujg}%
  \BibitemOpen
  \bibfield  {author} {\bibinfo {author} {\bibfnamefont {S.}~\bibnamefont
  {Profumo}}, \bibinfo {author} {\bibfnamefont {L.}~\bibnamefont {Giani}}, \
  and\ \bibinfo {author} {\bibfnamefont {O.~F.}\ \bibnamefont {Piattella}},\
  }\href {\doibase 10.3390/universe5100213} {\bibfield  {journal} {\bibinfo
  {journal} {Universe}\ }\textbf {\bibinfo {volume} {5}},\ \bibinfo {pages}
  {213} (\bibinfo {year} {2019})},\ \Eprint {http://arxiv.org/abs/1910.05610}
  {arXiv:1910.05610 [hep-ph]} \BibitemShut {NoStop}%
\bibitem [{\citenamefont {Camalich}\ and\ \citenamefont
  {Patel}(2022)}]{CAMALICH20221}%
  \BibitemOpen
  \bibfield  {author} {\bibinfo {author} {\bibfnamefont {J.~M.}\ \bibnamefont
  {Camalich}}\ and\ \bibinfo {author} {\bibfnamefont {M.}~\bibnamefont
  {Patel}},\ }\href {\doibase https://doi.org/10.1016/j.scib.2021.09.012}
  {\bibfield  {journal} {\bibinfo  {journal} {Science Bulletin}\ }\textbf
  {\bibinfo {volume} {67}},\ \bibinfo {pages} {1} (\bibinfo {year}
  {2022})}\BibitemShut {NoStop}%
\bibitem [{\citenamefont {Crivellin}\ and\ \citenamefont
  {Hoferichter}(2021)}]{Crivellin:2021sff}%
  \BibitemOpen
  \bibfield  {author} {\bibinfo {author} {\bibfnamefont {A.}~\bibnamefont
  {Crivellin}}\ and\ \bibinfo {author} {\bibfnamefont {M.}~\bibnamefont
  {Hoferichter}},\ }\href {\doibase 10.1126/science.abk2450} {\bibfield
  {journal} {\bibinfo  {journal} {Science}\ }\textbf {\bibinfo {volume}
  {374}},\ \bibinfo {pages} {1051} (\bibinfo {year} {2021})},\ \Eprint
  {http://arxiv.org/abs/2111.12739} {arXiv:2111.12739 [hep-ph]} \BibitemShut
  {NoStop}%
\bibitem [{\citenamefont {Fischer}\ \emph {et~al.}(2021)\citenamefont {Fischer}
  \emph {et~al.}}]{Fischer:2021sqw}%
  \BibitemOpen
  \bibfield  {author} {\bibinfo {author} {\bibfnamefont {O.}~\bibnamefont
  {Fischer}} \emph {et~al.},\ }\href@noop {} {\  (\bibinfo {year} {2021})},\
  \Eprint {http://arxiv.org/abs/2109.06065} {arXiv:2109.06065 [hep-ph]}
  \BibitemShut {NoStop}%
\bibitem [{\citenamefont {Arcadi}\ \emph {et~al.}(2018)\citenamefont {Arcadi},
  \citenamefont {Dutra}, \citenamefont {Ghosh}, \citenamefont {Lindner},
  \citenamefont {Mambrini}, \citenamefont {Pierre}, \citenamefont {Profumo},\
  and\ \citenamefont {Queiroz}}]{Arcadi:2017kky}%
  \BibitemOpen
  \bibfield  {author} {\bibinfo {author} {\bibfnamefont {G.}~\bibnamefont
  {Arcadi}}, \bibinfo {author} {\bibfnamefont {M.}~\bibnamefont {Dutra}},
  \bibinfo {author} {\bibfnamefont {P.}~\bibnamefont {Ghosh}}, \bibinfo
  {author} {\bibfnamefont {M.}~\bibnamefont {Lindner}}, \bibinfo {author}
  {\bibfnamefont {Y.}~\bibnamefont {Mambrini}}, \bibinfo {author}
  {\bibfnamefont {M.}~\bibnamefont {Pierre}}, \bibinfo {author} {\bibfnamefont
  {S.}~\bibnamefont {Profumo}}, \ and\ \bibinfo {author} {\bibfnamefont
  {F.~S.}\ \bibnamefont {Queiroz}},\ }\href {\doibase
  10.1140/epjc/s10052-018-5662-y} {\bibfield  {journal} {\bibinfo  {journal}
  {Eur. Phys. J. C}\ }\textbf {\bibinfo {volume} {78}},\ \bibinfo {pages} {203}
  (\bibinfo {year} {2018})},\ \Eprint {http://arxiv.org/abs/1703.07364}
  {arXiv:1703.07364 [hep-ph]} \BibitemShut {NoStop}%
\bibitem [{\citenamefont {Aaij}\ \emph {et~al.}(2013)\citenamefont {Aaij} \emph
  {et~al.}}]{LHCb:2013ghj}%
  \BibitemOpen
  \bibfield  {author} {\bibinfo {author} {\bibfnamefont {R.}~\bibnamefont
  {Aaij}} \emph {et~al.} (\bibinfo {collaboration} {LHCb}),\ }\href {\doibase
  10.1103/PhysRevLett.111.191801} {\bibfield  {journal} {\bibinfo  {journal}
  {Phys. Rev. Lett.}\ }\textbf {\bibinfo {volume} {111}},\ \bibinfo {pages}
  {191801} (\bibinfo {year} {2013})},\ \Eprint {http://arxiv.org/abs/1308.1707}
  {arXiv:1308.1707 [hep-ex]} \BibitemShut {NoStop}%
\bibitem [{\citenamefont {Aaij}\ \emph {et~al.}(2014)\citenamefont {Aaij} \emph
  {et~al.}}]{LHCb:2014vgu}%
  \BibitemOpen
  \bibfield  {author} {\bibinfo {author} {\bibfnamefont {R.}~\bibnamefont
  {Aaij}} \emph {et~al.} (\bibinfo {collaboration} {LHCb}),\ }\href {\doibase
  10.1103/PhysRevLett.113.151601} {\bibfield  {journal} {\bibinfo  {journal}
  {Phys. Rev. Lett.}\ }\textbf {\bibinfo {volume} {113}},\ \bibinfo {pages}
  {151601} (\bibinfo {year} {2014})},\ \Eprint {http://arxiv.org/abs/1406.6482}
  {arXiv:1406.6482 [hep-ex]} \BibitemShut {NoStop}%
\bibitem [{\citenamefont {Khachatryan}\ \emph {et~al.}(2015)\citenamefont
  {Khachatryan} \emph {et~al.}}]{CMS:2014xfa}%
  \BibitemOpen
  \bibfield  {author} {\bibinfo {author} {\bibfnamefont {V.}~\bibnamefont
  {Khachatryan}} \emph {et~al.} (\bibinfo {collaboration} {CMS, LHCb}),\ }\href
  {\doibase 10.1038/nature14474} {\bibfield  {journal} {\bibinfo  {journal}
  {Nature}\ }\textbf {\bibinfo {volume} {522}},\ \bibinfo {pages} {68}
  (\bibinfo {year} {2015})},\ \Eprint {http://arxiv.org/abs/1411.4413}
  {arXiv:1411.4413 [hep-ex]} \BibitemShut {NoStop}%
\bibitem [{\citenamefont {Aaij}\ \emph {et~al.}(2016)\citenamefont {Aaij} \emph
  {et~al.}}]{LHCb:2015svh}%
  \BibitemOpen
  \bibfield  {author} {\bibinfo {author} {\bibfnamefont {R.}~\bibnamefont
  {Aaij}} \emph {et~al.} (\bibinfo {collaboration} {LHCb}),\ }\href {\doibase
  10.1007/JHEP02(2016)104} {\bibfield  {journal} {\bibinfo  {journal} {JHEP}\
  }\textbf {\bibinfo {volume} {02}},\ \bibinfo {pages} {104} (\bibinfo {year}
  {2016})},\ \Eprint {http://arxiv.org/abs/1512.04442} {arXiv:1512.04442
  [hep-ex]} \BibitemShut {NoStop}%
\bibitem [{\citenamefont {Aaij}\ \emph {et~al.}(2017)\citenamefont {Aaij} \emph
  {et~al.}}]{LHCb:2017avl}%
  \BibitemOpen
  \bibfield  {author} {\bibinfo {author} {\bibfnamefont {R.}~\bibnamefont
  {Aaij}} \emph {et~al.} (\bibinfo {collaboration} {LHCb}),\ }\href {\doibase
  10.1007/JHEP08(2017)055} {\bibfield  {journal} {\bibinfo  {journal} {JHEP}\
  }\textbf {\bibinfo {volume} {08}},\ \bibinfo {pages} {055} (\bibinfo {year}
  {2017})},\ \Eprint {http://arxiv.org/abs/1705.05802} {arXiv:1705.05802
  [hep-ex]} \BibitemShut {NoStop}%
\bibitem [{\citenamefont {Aaboud}\ \emph {et~al.}(2019)\citenamefont {Aaboud}
  \emph {et~al.}}]{ATLAS:2018cur}%
  \BibitemOpen
  \bibfield  {author} {\bibinfo {author} {\bibfnamefont {M.}~\bibnamefont
  {Aaboud}} \emph {et~al.} (\bibinfo {collaboration} {ATLAS}),\ }\href
  {\doibase 10.1007/JHEP04(2019)098} {\bibfield  {journal} {\bibinfo  {journal}
  {JHEP}\ }\textbf {\bibinfo {volume} {04}},\ \bibinfo {pages} {098} (\bibinfo
  {year} {2019})},\ \Eprint {http://arxiv.org/abs/1812.03017} {arXiv:1812.03017
  [hep-ex]} \BibitemShut {NoStop}%
\bibitem [{\citenamefont {Sirunyan}\ \emph {et~al.}(2020)\citenamefont
  {Sirunyan} \emph {et~al.}}]{CMS:2019bbr}%
  \BibitemOpen
  \bibfield  {author} {\bibinfo {author} {\bibfnamefont {A.~M.}\ \bibnamefont
  {Sirunyan}} \emph {et~al.} (\bibinfo {collaboration} {CMS}),\ }\href
  {\doibase 10.1007/JHEP04(2020)188} {\bibfield  {journal} {\bibinfo  {journal}
  {JHEP}\ }\textbf {\bibinfo {volume} {04}},\ \bibinfo {pages} {188} (\bibinfo
  {year} {2020})},\ \Eprint {http://arxiv.org/abs/1910.12127} {arXiv:1910.12127
  [hep-ex]} \BibitemShut {NoStop}%
\bibitem [{\citenamefont {Aaij}\ \emph {et~al.}(2020)\citenamefont {Aaij} \emph
  {et~al.}}]{LHCb:2020lmf}%
  \BibitemOpen
  \bibfield  {author} {\bibinfo {author} {\bibfnamefont {R.}~\bibnamefont
  {Aaij}} \emph {et~al.} (\bibinfo {collaboration} {LHCb}),\ }\href {\doibase
  10.1103/PhysRevLett.125.011802} {\bibfield  {journal} {\bibinfo  {journal}
  {Phys. Rev. Lett.}\ }\textbf {\bibinfo {volume} {125}},\ \bibinfo {pages}
  {011802} (\bibinfo {year} {2020})},\ \Eprint
  {http://arxiv.org/abs/2003.04831} {arXiv:2003.04831 [hep-ex]} \BibitemShut
  {NoStop}%
\bibitem [{\citenamefont {Aaij}\ \emph {et~al.}(2021)\citenamefont {Aaij} \emph
  {et~al.}}]{LHCb:2020gog}%
  \BibitemOpen
  \bibfield  {author} {\bibinfo {author} {\bibfnamefont {R.}~\bibnamefont
  {Aaij}} \emph {et~al.} (\bibinfo {collaboration} {LHCb}),\ }\href {\doibase
  10.1103/PhysRevLett.126.161802} {\bibfield  {journal} {\bibinfo  {journal}
  {Phys. Rev. Lett.}\ }\textbf {\bibinfo {volume} {126}},\ \bibinfo {pages}
  {161802} (\bibinfo {year} {2021})},\ \Eprint
  {http://arxiv.org/abs/2012.13241} {arXiv:2012.13241 [hep-ex]} \BibitemShut
  {NoStop}%
\bibitem [{\citenamefont {Altmannshofer}\ and\ \citenamefont
  {Stangl}(2021)}]{Altmannshofer:2021qrr}%
  \BibitemOpen
  \bibfield  {author} {\bibinfo {author} {\bibfnamefont {W.}~\bibnamefont
  {Altmannshofer}}\ and\ \bibinfo {author} {\bibfnamefont {P.}~\bibnamefont
  {Stangl}},\ }\href {\doibase 10.1140/epjc/s10052-021-09725-1} {\bibfield
  {journal} {\bibinfo  {journal} {Eur. Phys. J. C}\ }\textbf {\bibinfo {volume}
  {81}},\ \bibinfo {pages} {952} (\bibinfo {year} {2021})},\ \Eprint
  {http://arxiv.org/abs/2103.13370} {arXiv:2103.13370 [hep-ph]} \BibitemShut
  {NoStop}%
\bibitem [{\citenamefont {Geng}\ \emph {et~al.}(2021)\citenamefont {Geng},
  \citenamefont {Grinstein}, \citenamefont {J\"ager}, \citenamefont {Li},
  \citenamefont {Martin~Camalich},\ and\ \citenamefont {Shi}}]{Geng:2021nhg}%
  \BibitemOpen
  \bibfield  {author} {\bibinfo {author} {\bibfnamefont {L.-S.}\ \bibnamefont
  {Geng}}, \bibinfo {author} {\bibfnamefont {B.}~\bibnamefont {Grinstein}},
  \bibinfo {author} {\bibfnamefont {S.}~\bibnamefont {J\"ager}}, \bibinfo
  {author} {\bibfnamefont {S.-Y.}\ \bibnamefont {Li}}, \bibinfo {author}
  {\bibfnamefont {J.}~\bibnamefont {Martin~Camalich}}, \ and\ \bibinfo {author}
  {\bibfnamefont {R.-X.}\ \bibnamefont {Shi}},\ }\href {\doibase
  10.1103/PhysRevD.104.035029} {\bibfield  {journal} {\bibinfo  {journal}
  {Phys. Rev. D}\ }\textbf {\bibinfo {volume} {104}},\ \bibinfo {pages}
  {035029} (\bibinfo {year} {2021})},\ \Eprint
  {http://arxiv.org/abs/2103.12738} {arXiv:2103.12738 [hep-ph]} \BibitemShut
  {NoStop}%
\bibitem [{\citenamefont {Alguer\'o}\ \emph {et~al.}(2022)\citenamefont
  {Alguer\'o}, \citenamefont {Capdevila}, \citenamefont {Descotes-Genon},
  \citenamefont {Matias},\ and\ \citenamefont
  {Novoa-Brunet}}]{Alguero:2021anc}%
  \BibitemOpen
  \bibfield  {author} {\bibinfo {author} {\bibfnamefont {M.}~\bibnamefont
  {Alguer\'o}}, \bibinfo {author} {\bibfnamefont {B.}~\bibnamefont
  {Capdevila}}, \bibinfo {author} {\bibfnamefont {S.}~\bibnamefont
  {Descotes-Genon}}, \bibinfo {author} {\bibfnamefont {J.}~\bibnamefont
  {Matias}}, \ and\ \bibinfo {author} {\bibfnamefont {M.}~\bibnamefont
  {Novoa-Brunet}},\ }\href {\doibase 10.1140/epjc/s10052-022-10231-1}
  {\bibfield  {journal} {\bibinfo  {journal} {Eur. Phys. J. C}\ }\textbf
  {\bibinfo {volume} {82}},\ \bibinfo {pages} {326} (\bibinfo {year} {2022})},\
  \Eprint {http://arxiv.org/abs/2104.08921} {arXiv:2104.08921 [hep-ph]}
  \BibitemShut {NoStop}%
\bibitem [{\citenamefont {Hurth}\ \emph {et~al.}(2022)\citenamefont {Hurth},
  \citenamefont {Mahmoudi}, \citenamefont {Santos},\ and\ \citenamefont
  {Neshatpour}}]{Hurth:2021nsi}%
  \BibitemOpen
  \bibfield  {author} {\bibinfo {author} {\bibfnamefont {T.}~\bibnamefont
  {Hurth}}, \bibinfo {author} {\bibfnamefont {F.}~\bibnamefont {Mahmoudi}},
  \bibinfo {author} {\bibfnamefont {D.~M.}\ \bibnamefont {Santos}}, \ and\
  \bibinfo {author} {\bibfnamefont {S.}~\bibnamefont {Neshatpour}},\ }\href
  {\doibase 10.1016/j.physletb.2021.136838} {\bibfield  {journal} {\bibinfo
  {journal} {Phys. Lett. B}\ }\textbf {\bibinfo {volume} {824}},\ \bibinfo
  {pages} {136838} (\bibinfo {year} {2022})},\ \Eprint
  {http://arxiv.org/abs/2104.10058} {arXiv:2104.10058 [hep-ph]} \BibitemShut
  {NoStop}%
\bibitem [{\citenamefont {Isidori}\ \emph {et~al.}(2021)\citenamefont
  {Isidori}, \citenamefont {Lancierini}, \citenamefont {Owen},\ and\
  \citenamefont {Serra}}]{Isidori:2021vtc}%
  \BibitemOpen
  \bibfield  {author} {\bibinfo {author} {\bibfnamefont {G.}~\bibnamefont
  {Isidori}}, \bibinfo {author} {\bibfnamefont {D.}~\bibnamefont {Lancierini}},
  \bibinfo {author} {\bibfnamefont {P.}~\bibnamefont {Owen}}, \ and\ \bibinfo
  {author} {\bibfnamefont {N.}~\bibnamefont {Serra}},\ }\href {\doibase
  10.1016/j.physletb.2021.136644} {\bibfield  {journal} {\bibinfo  {journal}
  {Phys. Lett. B}\ }\textbf {\bibinfo {volume} {822}},\ \bibinfo {pages}
  {136644} (\bibinfo {year} {2021})},\ \Eprint
  {http://arxiv.org/abs/2104.05631} {arXiv:2104.05631 [hep-ph]} \BibitemShut
  {NoStop}%
\bibitem [{\citenamefont {Ciuchini}\ \emph
  {et~al.}(2021{\natexlab{a}})\citenamefont {Ciuchini}, \citenamefont {Fedele},
  \citenamefont {Franco}, \citenamefont {Paul}, \citenamefont {Silvestrini},\
  and\ \citenamefont {Valli}}]{Ciuchini:2021smi}%
  \BibitemOpen
  \bibfield  {author} {\bibinfo {author} {\bibfnamefont {M.}~\bibnamefont
  {Ciuchini}}, \bibinfo {author} {\bibfnamefont {M.}~\bibnamefont {Fedele}},
  \bibinfo {author} {\bibfnamefont {E.}~\bibnamefont {Franco}}, \bibinfo
  {author} {\bibfnamefont {A.}~\bibnamefont {Paul}}, \bibinfo {author}
  {\bibfnamefont {L.}~\bibnamefont {Silvestrini}}, \ and\ \bibinfo {author}
  {\bibfnamefont {M.}~\bibnamefont {Valli}},\ }\href@noop {} {\  (\bibinfo
  {year} {2021}{\natexlab{a}})},\ \Eprint {http://arxiv.org/abs/2110.10126}
  {arXiv:2110.10126 [hep-ph]} \BibitemShut {NoStop}%
\bibitem [{\citenamefont {Alonso}\ \emph {et~al.}(2015)\citenamefont {Alonso},
  \citenamefont {Grinstein},\ and\ \citenamefont
  {Martin~Camalich}}]{Alonso:2015sja}%
  \BibitemOpen
  \bibfield  {author} {\bibinfo {author} {\bibfnamefont {R.}~\bibnamefont
  {Alonso}}, \bibinfo {author} {\bibfnamefont {B.}~\bibnamefont {Grinstein}}, \
  and\ \bibinfo {author} {\bibfnamefont {J.}~\bibnamefont {Martin~Camalich}},\
  }\href {\doibase 10.1007/JHEP10(2015)184} {\bibfield  {journal} {\bibinfo
  {journal} {JHEP}\ }\textbf {\bibinfo {volume} {10}},\ \bibinfo {pages} {184}
  (\bibinfo {year} {2015})},\ \Eprint {http://arxiv.org/abs/1505.05164}
  {arXiv:1505.05164 [hep-ph]} \BibitemShut {NoStop}%
\bibitem [{\citenamefont {Calibbi}\ \emph {et~al.}(2015)\citenamefont
  {Calibbi}, \citenamefont {Crivellin},\ and\ \citenamefont
  {Ota}}]{Calibbi:2015kma}%
  \BibitemOpen
  \bibfield  {author} {\bibinfo {author} {\bibfnamefont {L.}~\bibnamefont
  {Calibbi}}, \bibinfo {author} {\bibfnamefont {A.}~\bibnamefont {Crivellin}},
  \ and\ \bibinfo {author} {\bibfnamefont {T.}~\bibnamefont {Ota}},\ }\href
  {\doibase 10.1103/PhysRevLett.115.181801} {\bibfield  {journal} {\bibinfo
  {journal} {Phys. Rev. Lett.}\ }\textbf {\bibinfo {volume} {115}},\ \bibinfo
  {pages} {181801} (\bibinfo {year} {2015})},\ \Eprint
  {http://arxiv.org/abs/1506.02661} {arXiv:1506.02661 [hep-ph]} \BibitemShut
  {NoStop}%
\bibitem [{\citenamefont {Hiller}\ \emph {et~al.}(2016)\citenamefont {Hiller},
  \citenamefont {Loose},\ and\ \citenamefont {Sch\"onwald}}]{Hiller:2016kry}%
  \BibitemOpen
  \bibfield  {author} {\bibinfo {author} {\bibfnamefont {G.}~\bibnamefont
  {Hiller}}, \bibinfo {author} {\bibfnamefont {D.}~\bibnamefont {Loose}}, \
  and\ \bibinfo {author} {\bibfnamefont {K.}~\bibnamefont {Sch\"onwald}},\
  }\href {\doibase 10.1007/JHEP12(2016)027} {\bibfield  {journal} {\bibinfo
  {journal} {JHEP}\ }\textbf {\bibinfo {volume} {12}},\ \bibinfo {pages} {027}
  (\bibinfo {year} {2016})},\ \Eprint {http://arxiv.org/abs/1609.08895}
  {arXiv:1609.08895 [hep-ph]} \BibitemShut {NoStop}%
\bibitem [{\citenamefont {Bhattacharya}\ \emph {et~al.}(2017)\citenamefont
  {Bhattacharya}, \citenamefont {Datta}, \citenamefont {Gu\'evin},
  \citenamefont {London},\ and\ \citenamefont
  {Watanabe}}]{Bhattacharya:2016mcc}%
  \BibitemOpen
  \bibfield  {author} {\bibinfo {author} {\bibfnamefont {B.}~\bibnamefont
  {Bhattacharya}}, \bibinfo {author} {\bibfnamefont {A.}~\bibnamefont {Datta}},
  \bibinfo {author} {\bibfnamefont {J.-P.}\ \bibnamefont {Gu\'evin}}, \bibinfo
  {author} {\bibfnamefont {D.}~\bibnamefont {London}}, \ and\ \bibinfo {author}
  {\bibfnamefont {R.}~\bibnamefont {Watanabe}},\ }\href {\doibase
  10.1007/JHEP01(2017)015} {\bibfield  {journal} {\bibinfo  {journal} {JHEP}\
  }\textbf {\bibinfo {volume} {01}},\ \bibinfo {pages} {015} (\bibinfo {year}
  {2017})},\ \Eprint {http://arxiv.org/abs/1609.09078} {arXiv:1609.09078
  [hep-ph]} \BibitemShut {NoStop}%
\bibitem [{\citenamefont {Buttazzo}\ \emph {et~al.}(2017)\citenamefont
  {Buttazzo}, \citenamefont {Greljo}, \citenamefont {Isidori},\ and\
  \citenamefont {Marzocca}}]{Buttazzo:2017ixm}%
  \BibitemOpen
  \bibfield  {author} {\bibinfo {author} {\bibfnamefont {D.}~\bibnamefont
  {Buttazzo}}, \bibinfo {author} {\bibfnamefont {A.}~\bibnamefont {Greljo}},
  \bibinfo {author} {\bibfnamefont {G.}~\bibnamefont {Isidori}}, \ and\
  \bibinfo {author} {\bibfnamefont {D.}~\bibnamefont {Marzocca}},\ }\href
  {\doibase 10.1007/JHEP11(2017)044} {\bibfield  {journal} {\bibinfo  {journal}
  {JHEP}\ }\textbf {\bibinfo {volume} {11}},\ \bibinfo {pages} {044} (\bibinfo
  {year} {2017})},\ \Eprint {http://arxiv.org/abs/1706.07808} {arXiv:1706.07808
  [hep-ph]} \BibitemShut {NoStop}%
\bibitem [{\citenamefont {Barbieri}\ \emph {et~al.}(2016)\citenamefont
  {Barbieri}, \citenamefont {Isidori}, \citenamefont {Pattori},\ and\
  \citenamefont {Senia}}]{Barbieri:2015yvd}%
  \BibitemOpen
  \bibfield  {author} {\bibinfo {author} {\bibfnamefont {R.}~\bibnamefont
  {Barbieri}}, \bibinfo {author} {\bibfnamefont {G.}~\bibnamefont {Isidori}},
  \bibinfo {author} {\bibfnamefont {A.}~\bibnamefont {Pattori}}, \ and\
  \bibinfo {author} {\bibfnamefont {F.}~\bibnamefont {Senia}},\ }\href
  {\doibase 10.1140/epjc/s10052-016-3905-3} {\bibfield  {journal} {\bibinfo
  {journal} {Eur. Phys. J. C}\ }\textbf {\bibinfo {volume} {76}},\ \bibinfo
  {pages} {67} (\bibinfo {year} {2016})},\ \Eprint
  {http://arxiv.org/abs/1512.01560} {arXiv:1512.01560 [hep-ph]} \BibitemShut
  {NoStop}%
\bibitem [{\citenamefont {Barbieri}\ \emph {et~al.}(2017)\citenamefont
  {Barbieri}, \citenamefont {Murphy},\ and\ \citenamefont
  {Senia}}]{Barbieri:2016las}%
  \BibitemOpen
  \bibfield  {author} {\bibinfo {author} {\bibfnamefont {R.}~\bibnamefont
  {Barbieri}}, \bibinfo {author} {\bibfnamefont {C.~W.}\ \bibnamefont
  {Murphy}}, \ and\ \bibinfo {author} {\bibfnamefont {F.}~\bibnamefont
  {Senia}},\ }\href {\doibase 10.1140/epjc/s10052-016-4578-7} {\bibfield
  {journal} {\bibinfo  {journal} {Eur. Phys. J. C}\ }\textbf {\bibinfo {volume}
  {77}},\ \bibinfo {pages} {8} (\bibinfo {year} {2017})},\ \Eprint
  {http://arxiv.org/abs/1611.04930} {arXiv:1611.04930 [hep-ph]} \BibitemShut
  {NoStop}%
\bibitem [{\citenamefont {Calibbi}\ \emph {et~al.}(2018)\citenamefont
  {Calibbi}, \citenamefont {Crivellin},\ and\ \citenamefont
  {Li}}]{Calibbi:2017qbu}%
  \BibitemOpen
  \bibfield  {author} {\bibinfo {author} {\bibfnamefont {L.}~\bibnamefont
  {Calibbi}}, \bibinfo {author} {\bibfnamefont {A.}~\bibnamefont {Crivellin}},
  \ and\ \bibinfo {author} {\bibfnamefont {T.}~\bibnamefont {Li}},\ }\href
  {\doibase 10.1103/PhysRevD.98.115002} {\bibfield  {journal} {\bibinfo
  {journal} {Phys. Rev. D}\ }\textbf {\bibinfo {volume} {98}},\ \bibinfo
  {pages} {115002} (\bibinfo {year} {2018})},\ \Eprint
  {http://arxiv.org/abs/1709.00692} {arXiv:1709.00692 [hep-ph]} \BibitemShut
  {NoStop}%
\bibitem [{\citenamefont {Crivellin}\ \emph {et~al.}(2017)\citenamefont
  {Crivellin}, \citenamefont {M\"uller},\ and\ \citenamefont
  {Ota}}]{Crivellin:2017zlb}%
  \BibitemOpen
  \bibfield  {author} {\bibinfo {author} {\bibfnamefont {A.}~\bibnamefont
  {Crivellin}}, \bibinfo {author} {\bibfnamefont {D.}~\bibnamefont {M\"uller}},
  \ and\ \bibinfo {author} {\bibfnamefont {T.}~\bibnamefont {Ota}},\ }\href
  {\doibase 10.1007/JHEP09(2017)040} {\bibfield  {journal} {\bibinfo  {journal}
  {JHEP}\ }\textbf {\bibinfo {volume} {09}},\ \bibinfo {pages} {040} (\bibinfo
  {year} {2017})},\ \Eprint {http://arxiv.org/abs/1703.09226} {arXiv:1703.09226
  [hep-ph]} \BibitemShut {NoStop}%
\bibitem [{\citenamefont {Crivellin}\ \emph {et~al.}(2018)\citenamefont
  {Crivellin}, \citenamefont {M\"uller}, \citenamefont {Signer},\ and\
  \citenamefont {Ulrich}}]{Crivellin:2017dsk}%
  \BibitemOpen
  \bibfield  {author} {\bibinfo {author} {\bibfnamefont {A.}~\bibnamefont
  {Crivellin}}, \bibinfo {author} {\bibfnamefont {D.}~\bibnamefont {M\"uller}},
  \bibinfo {author} {\bibfnamefont {A.}~\bibnamefont {Signer}}, \ and\ \bibinfo
  {author} {\bibfnamefont {Y.}~\bibnamefont {Ulrich}},\ }\href {\doibase
  10.1103/PhysRevD.97.015019} {\bibfield  {journal} {\bibinfo  {journal} {Phys.
  Rev. D}\ }\textbf {\bibinfo {volume} {97}},\ \bibinfo {pages} {015019}
  (\bibinfo {year} {2018})},\ \Eprint {http://arxiv.org/abs/1706.08511}
  {arXiv:1706.08511 [hep-ph]} \BibitemShut {NoStop}%
\bibitem [{\citenamefont {Bordone}\ \emph {et~al.}(2018)\citenamefont
  {Bordone}, \citenamefont {Cornella}, \citenamefont {Fuentes-Mart\'\i{}n},\
  and\ \citenamefont {Isidori}}]{Bordone:2018nbg}%
  \BibitemOpen
  \bibfield  {author} {\bibinfo {author} {\bibfnamefont {M.}~\bibnamefont
  {Bordone}}, \bibinfo {author} {\bibfnamefont {C.}~\bibnamefont {Cornella}},
  \bibinfo {author} {\bibfnamefont {J.}~\bibnamefont {Fuentes-Mart\'\i{}n}}, \
  and\ \bibinfo {author} {\bibfnamefont {G.}~\bibnamefont {Isidori}},\ }\href
  {\doibase 10.1007/JHEP10(2018)148} {\bibfield  {journal} {\bibinfo  {journal}
  {JHEP}\ }\textbf {\bibinfo {volume} {10}},\ \bibinfo {pages} {148} (\bibinfo
  {year} {2018})},\ \Eprint {http://arxiv.org/abs/1805.09328} {arXiv:1805.09328
  [hep-ph]} \BibitemShut {NoStop}%
\bibitem [{\citenamefont {Kumar}\ \emph {et~al.}(2019)\citenamefont {Kumar},
  \citenamefont {London},\ and\ \citenamefont {Watanabe}}]{Kumar:2018kmr}%
  \BibitemOpen
  \bibfield  {author} {\bibinfo {author} {\bibfnamefont {J.}~\bibnamefont
  {Kumar}}, \bibinfo {author} {\bibfnamefont {D.}~\bibnamefont {London}}, \
  and\ \bibinfo {author} {\bibfnamefont {R.}~\bibnamefont {Watanabe}},\ }\href
  {\doibase 10.1103/PhysRevD.99.015007} {\bibfield  {journal} {\bibinfo
  {journal} {Phys. Rev. D}\ }\textbf {\bibinfo {volume} {99}},\ \bibinfo
  {pages} {015007} (\bibinfo {year} {2019})},\ \Eprint
  {http://arxiv.org/abs/1806.07403} {arXiv:1806.07403 [hep-ph]} \BibitemShut
  {NoStop}%
\bibitem [{\citenamefont {Crivellin}\ \emph {et~al.}(2019)\citenamefont
  {Crivellin}, \citenamefont {Greub}, \citenamefont {M\"uller},\ and\
  \citenamefont {Saturnino}}]{Crivellin:2018yvo}%
  \BibitemOpen
  \bibfield  {author} {\bibinfo {author} {\bibfnamefont {A.}~\bibnamefont
  {Crivellin}}, \bibinfo {author} {\bibfnamefont {C.}~\bibnamefont {Greub}},
  \bibinfo {author} {\bibfnamefont {D.}~\bibnamefont {M\"uller}}, \ and\
  \bibinfo {author} {\bibfnamefont {F.}~\bibnamefont {Saturnino}},\ }\href
  {\doibase 10.1103/PhysRevLett.122.011805} {\bibfield  {journal} {\bibinfo
  {journal} {Phys. Rev. Lett.}\ }\textbf {\bibinfo {volume} {122}},\ \bibinfo
  {pages} {011805} (\bibinfo {year} {2019})},\ \Eprint
  {http://arxiv.org/abs/1807.02068} {arXiv:1807.02068 [hep-ph]} \BibitemShut
  {NoStop}%
\bibitem [{\citenamefont {Crivellin}\ and\ \citenamefont
  {Saturnino}(2019)}]{Crivellin:2019szf}%
  \BibitemOpen
  \bibfield  {author} {\bibinfo {author} {\bibfnamefont {A.}~\bibnamefont
  {Crivellin}}\ and\ \bibinfo {author} {\bibfnamefont {F.}~\bibnamefont
  {Saturnino}},\ }\href {\doibase 10.22323/1.352.0163} {\bibfield  {journal}
  {\bibinfo  {journal} {PoS}\ }\textbf {\bibinfo {volume} {DIS2019}},\ \bibinfo
  {pages} {163} (\bibinfo {year} {2019})},\ \Eprint
  {http://arxiv.org/abs/1906.01222} {arXiv:1906.01222 [hep-ph]} \BibitemShut
  {NoStop}%
\bibitem [{\citenamefont {Cornella}\ \emph {et~al.}(2019)\citenamefont
  {Cornella}, \citenamefont {Fuentes-Martin},\ and\ \citenamefont
  {Isidori}}]{Cornella:2019hct}%
  \BibitemOpen
  \bibfield  {author} {\bibinfo {author} {\bibfnamefont {C.}~\bibnamefont
  {Cornella}}, \bibinfo {author} {\bibfnamefont {J.}~\bibnamefont
  {Fuentes-Martin}}, \ and\ \bibinfo {author} {\bibfnamefont {G.}~\bibnamefont
  {Isidori}},\ }\href {\doibase 10.1007/JHEP07(2019)168} {\bibfield  {journal}
  {\bibinfo  {journal} {JHEP}\ }\textbf {\bibinfo {volume} {07}},\ \bibinfo
  {pages} {168} (\bibinfo {year} {2019})},\ \Eprint
  {http://arxiv.org/abs/1903.11517} {arXiv:1903.11517 [hep-ph]} \BibitemShut
  {NoStop}%
\bibitem [{\citenamefont {Bordone}\ \emph {et~al.}(2020)\citenamefont
  {Bordone}, \citenamefont {Cat\`a},\ and\ \citenamefont
  {Feldmann}}]{Bordone:2019uzc}%
  \BibitemOpen
  \bibfield  {author} {\bibinfo {author} {\bibfnamefont {M.}~\bibnamefont
  {Bordone}}, \bibinfo {author} {\bibfnamefont {O.}~\bibnamefont {Cat\`a}}, \
  and\ \bibinfo {author} {\bibfnamefont {T.}~\bibnamefont {Feldmann}},\ }\href
  {\doibase 10.1007/JHEP01(2020)067} {\bibfield  {journal} {\bibinfo  {journal}
  {JHEP}\ }\textbf {\bibinfo {volume} {01}},\ \bibinfo {pages} {067} (\bibinfo
  {year} {2020})},\ \Eprint {http://arxiv.org/abs/1910.02641} {arXiv:1910.02641
  [hep-ph]} \BibitemShut {NoStop}%
\bibitem [{\citenamefont {Bernigaud}\ \emph {et~al.}(2020)\citenamefont
  {Bernigaud}, \citenamefont {de~Medeiros~Varzielas},\ and\ \citenamefont
  {Talbert}}]{Bernigaud:2019bfy}%
  \BibitemOpen
  \bibfield  {author} {\bibinfo {author} {\bibfnamefont {J.}~\bibnamefont
  {Bernigaud}}, \bibinfo {author} {\bibfnamefont {I.}~\bibnamefont
  {de~Medeiros~Varzielas}}, \ and\ \bibinfo {author} {\bibfnamefont
  {J.}~\bibnamefont {Talbert}},\ }\href {\doibase 10.1007/JHEP01(2020)194}
  {\bibfield  {journal} {\bibinfo  {journal} {JHEP}\ }\textbf {\bibinfo
  {volume} {01}},\ \bibinfo {pages} {194} (\bibinfo {year} {2020})},\ \Eprint
  {http://arxiv.org/abs/1906.11270} {arXiv:1906.11270 [hep-ph]} \BibitemShut
  {NoStop}%
\bibitem [{\citenamefont {Aebischer}\ \emph {et~al.}(2019)\citenamefont
  {Aebischer}, \citenamefont {Crivellin},\ and\ \citenamefont
  {Greub}}]{Aebischer:2018acj}%
  \BibitemOpen
  \bibfield  {author} {\bibinfo {author} {\bibfnamefont {J.}~\bibnamefont
  {Aebischer}}, \bibinfo {author} {\bibfnamefont {A.}~\bibnamefont
  {Crivellin}}, \ and\ \bibinfo {author} {\bibfnamefont {C.}~\bibnamefont
  {Greub}},\ }\href {\doibase 10.1103/PhysRevD.99.055002} {\bibfield  {journal}
  {\bibinfo  {journal} {Phys. Rev. D}\ }\textbf {\bibinfo {volume} {99}},\
  \bibinfo {pages} {055002} (\bibinfo {year} {2019})},\ \Eprint
  {http://arxiv.org/abs/1811.08907} {arXiv:1811.08907 [hep-ph]} \BibitemShut
  {NoStop}%
\bibitem [{\citenamefont {Fuentes-Mart\'\i{}n}\ \emph
  {et~al.}(2020)\citenamefont {Fuentes-Mart\'\i{}n}, \citenamefont {Isidori},
  \citenamefont {K\"onig},\ and\ \citenamefont
  {Selimovi\'c}}]{Fuentes-Martin:2019ign}%
  \BibitemOpen
  \bibfield  {author} {\bibinfo {author} {\bibfnamefont {J.}~\bibnamefont
  {Fuentes-Mart\'\i{}n}}, \bibinfo {author} {\bibfnamefont {G.}~\bibnamefont
  {Isidori}}, \bibinfo {author} {\bibfnamefont {M.}~\bibnamefont {K\"onig}}, \
  and\ \bibinfo {author} {\bibfnamefont {N.}~\bibnamefont {Selimovi\'c}},\
  }\href {\doibase 10.1103/PhysRevD.101.035024} {\bibfield  {journal} {\bibinfo
   {journal} {Phys. Rev. D}\ }\textbf {\bibinfo {volume} {101}},\ \bibinfo
  {pages} {035024} (\bibinfo {year} {2020})},\ \Eprint
  {http://arxiv.org/abs/1910.13474} {arXiv:1910.13474 [hep-ph]} \BibitemShut
  {NoStop}%
\bibitem [{\citenamefont {Popov}\ \emph {et~al.}(2019)\citenamefont {Popov},
  \citenamefont {Schmidt},\ and\ \citenamefont {White}}]{Popov:2019tyc}%
  \BibitemOpen
  \bibfield  {author} {\bibinfo {author} {\bibfnamefont {O.}~\bibnamefont
  {Popov}}, \bibinfo {author} {\bibfnamefont {M.~A.}\ \bibnamefont {Schmidt}},
  \ and\ \bibinfo {author} {\bibfnamefont {G.}~\bibnamefont {White}},\ }\href
  {\doibase 10.1103/PhysRevD.100.035028} {\bibfield  {journal} {\bibinfo
  {journal} {Phys. Rev. D}\ }\textbf {\bibinfo {volume} {100}},\ \bibinfo
  {pages} {035028} (\bibinfo {year} {2019})},\ \Eprint
  {http://arxiv.org/abs/1905.06339} {arXiv:1905.06339 [hep-ph]} \BibitemShut
  {NoStop}%
\bibitem [{\citenamefont {Fajfer}\ and\ \citenamefont
  {Ko\v{s}nik}(2016)}]{Fajfer:2015ycq}%
  \BibitemOpen
  \bibfield  {author} {\bibinfo {author} {\bibfnamefont {S.}~\bibnamefont
  {Fajfer}}\ and\ \bibinfo {author} {\bibfnamefont {N.}~\bibnamefont
  {Ko\v{s}nik}},\ }\href {\doibase 10.1016/j.physletb.2016.02.018} {\bibfield
  {journal} {\bibinfo  {journal} {Phys. Lett. B}\ }\textbf {\bibinfo {volume}
  {755}},\ \bibinfo {pages} {270} (\bibinfo {year} {2016})},\ \Eprint
  {http://arxiv.org/abs/1511.06024} {arXiv:1511.06024 [hep-ph]} \BibitemShut
  {NoStop}%
\bibitem [{\citenamefont {Blanke}\ and\ \citenamefont
  {Crivellin}(2018)}]{Blanke:2018sro}%
  \BibitemOpen
  \bibfield  {author} {\bibinfo {author} {\bibfnamefont {M.}~\bibnamefont
  {Blanke}}\ and\ \bibinfo {author} {\bibfnamefont {A.}~\bibnamefont
  {Crivellin}},\ }\href {\doibase 10.1103/PhysRevLett.121.011801} {\bibfield
  {journal} {\bibinfo  {journal} {Phys. Rev. Lett.}\ }\textbf {\bibinfo
  {volume} {121}},\ \bibinfo {pages} {011801} (\bibinfo {year} {2018})},\
  \Eprint {http://arxiv.org/abs/1801.07256} {arXiv:1801.07256 [hep-ph]}
  \BibitemShut {NoStop}%
\bibitem [{\citenamefont {de~Medeiros~Varzielas}\ and\ \citenamefont
  {Talbert}(2019)}]{deMedeirosVarzielas:2019lgb}%
  \BibitemOpen
  \bibfield  {author} {\bibinfo {author} {\bibfnamefont {I.}~\bibnamefont
  {de~Medeiros~Varzielas}}\ and\ \bibinfo {author} {\bibfnamefont
  {J.}~\bibnamefont {Talbert}},\ }\href {\doibase
  10.1140/epjc/s10052-019-7047-2} {\bibfield  {journal} {\bibinfo  {journal}
  {Eur. Phys. J. C}\ }\textbf {\bibinfo {volume} {79}},\ \bibinfo {pages} {536}
  (\bibinfo {year} {2019})},\ \Eprint {http://arxiv.org/abs/1901.10484}
  {arXiv:1901.10484 [hep-ph]} \BibitemShut {NoStop}%
\bibitem [{\citenamefont {de~Medeiros~Varzielas}\ and\ \citenamefont
  {Hiller}(2015)}]{Varzielas:2015iva}%
  \BibitemOpen
  \bibfield  {author} {\bibinfo {author} {\bibfnamefont {I.}~\bibnamefont
  {de~Medeiros~Varzielas}}\ and\ \bibinfo {author} {\bibfnamefont
  {G.}~\bibnamefont {Hiller}},\ }\href {\doibase 10.1007/JHEP06(2015)072}
  {\bibfield  {journal} {\bibinfo  {journal} {JHEP}\ }\textbf {\bibinfo
  {volume} {06}},\ \bibinfo {pages} {072} (\bibinfo {year} {2015})},\ \Eprint
  {http://arxiv.org/abs/1503.01084} {arXiv:1503.01084 [hep-ph]} \BibitemShut
  {NoStop}%
\bibitem [{\citenamefont {Crivellin}\ \emph {et~al.}(2020)\citenamefont
  {Crivellin}, \citenamefont {M\"uller},\ and\ \citenamefont
  {Saturnino}}]{Crivellin:2019dwb}%
  \BibitemOpen
  \bibfield  {author} {\bibinfo {author} {\bibfnamefont {A.}~\bibnamefont
  {Crivellin}}, \bibinfo {author} {\bibfnamefont {D.}~\bibnamefont {M\"uller}},
  \ and\ \bibinfo {author} {\bibfnamefont {F.}~\bibnamefont {Saturnino}},\
  }\href {\doibase 10.1007/JHEP06(2020)020} {\bibfield  {journal} {\bibinfo
  {journal} {JHEP}\ }\textbf {\bibinfo {volume} {06}},\ \bibinfo {pages} {020}
  (\bibinfo {year} {2020})},\ \Eprint {http://arxiv.org/abs/1912.04224}
  {arXiv:1912.04224 [hep-ph]} \BibitemShut {NoStop}%
\bibitem [{\citenamefont {Saad}(2020)}]{Saad:2020ihm}%
  \BibitemOpen
  \bibfield  {author} {\bibinfo {author} {\bibfnamefont {S.}~\bibnamefont
  {Saad}},\ }\href {\doibase 10.1103/PhysRevD.102.015019} {\bibfield  {journal}
  {\bibinfo  {journal} {Phys. Rev. D}\ }\textbf {\bibinfo {volume} {102}},\
  \bibinfo {pages} {015019} (\bibinfo {year} {2020})},\ \Eprint
  {http://arxiv.org/abs/2005.04352} {arXiv:2005.04352 [hep-ph]} \BibitemShut
  {NoStop}%
\bibitem [{\citenamefont {Saad}\ and\ \citenamefont
  {Thapa}(2020)}]{Saad:2020ucl}%
  \BibitemOpen
  \bibfield  {author} {\bibinfo {author} {\bibfnamefont {S.}~\bibnamefont
  {Saad}}\ and\ \bibinfo {author} {\bibfnamefont {A.}~\bibnamefont {Thapa}},\
  }\href {\doibase 10.1103/PhysRevD.102.015014} {\bibfield  {journal} {\bibinfo
   {journal} {Phys. Rev. D}\ }\textbf {\bibinfo {volume} {102}},\ \bibinfo
  {pages} {015014} (\bibinfo {year} {2020})},\ \Eprint
  {http://arxiv.org/abs/2004.07880} {arXiv:2004.07880 [hep-ph]} \BibitemShut
  {NoStop}%
\bibitem [{\citenamefont {Gherardi}\ \emph {et~al.}(2021)\citenamefont
  {Gherardi}, \citenamefont {Marzocca},\ and\ \citenamefont
  {Venturini}}]{Gherardi:2020qhc}%
  \BibitemOpen
  \bibfield  {author} {\bibinfo {author} {\bibfnamefont {V.}~\bibnamefont
  {Gherardi}}, \bibinfo {author} {\bibfnamefont {D.}~\bibnamefont {Marzocca}},
  \ and\ \bibinfo {author} {\bibfnamefont {E.}~\bibnamefont {Venturini}},\
  }\href {\doibase 10.1007/JHEP01(2021)138} {\bibfield  {journal} {\bibinfo
  {journal} {JHEP}\ }\textbf {\bibinfo {volume} {01}},\ \bibinfo {pages} {138}
  (\bibinfo {year} {2021})},\ \Eprint {http://arxiv.org/abs/2008.09548}
  {arXiv:2008.09548 [hep-ph]} \BibitemShut {NoStop}%
\bibitem [{\citenamefont {Da~Rold}\ and\ \citenamefont
  {Lamagna}(2021)}]{DaRold:2020bib}%
  \BibitemOpen
  \bibfield  {author} {\bibinfo {author} {\bibfnamefont {L.}~\bibnamefont
  {Da~Rold}}\ and\ \bibinfo {author} {\bibfnamefont {F.}~\bibnamefont
  {Lamagna}},\ }\href {\doibase 10.1103/PhysRevD.103.115007} {\bibfield
  {journal} {\bibinfo  {journal} {Phys. Rev. D}\ }\textbf {\bibinfo {volume}
  {103}},\ \bibinfo {pages} {115007} (\bibinfo {year} {2021})},\ \Eprint
  {http://arxiv.org/abs/2011.10061} {arXiv:2011.10061 [hep-ph]} \BibitemShut
  {NoStop}%
\bibitem [{\citenamefont {Heeck}\ and\ \citenamefont
  {Thapa}(2022)}]{Heeck:2022znj}%
  \BibitemOpen
  \bibfield  {author} {\bibinfo {author} {\bibfnamefont {J.}~\bibnamefont
  {Heeck}}\ and\ \bibinfo {author} {\bibfnamefont {A.}~\bibnamefont {Thapa}},\
  }\href@noop {} {\  (\bibinfo {year} {2022})},\ \Eprint
  {http://arxiv.org/abs/2202.08854} {arXiv:2202.08854 [hep-ph]} \BibitemShut
  {NoStop}%
\bibitem [{\citenamefont {Choi}\ \emph {et~al.}(2018)\citenamefont {Choi},
  \citenamefont {Kang}, \citenamefont {Lee},\ and\ \citenamefont
  {Ro}}]{Choi:2018stw}%
  \BibitemOpen
  \bibfield  {author} {\bibinfo {author} {\bibfnamefont {S.-M.}\ \bibnamefont
  {Choi}}, \bibinfo {author} {\bibfnamefont {Y.-J.}\ \bibnamefont {Kang}},
  \bibinfo {author} {\bibfnamefont {H.~M.}\ \bibnamefont {Lee}}, \ and\
  \bibinfo {author} {\bibfnamefont {T.-G.}\ \bibnamefont {Ro}},\ }\href
  {\doibase 10.1007/JHEP10(2018)104} {\bibfield  {journal} {\bibinfo  {journal}
  {JHEP}\ }\textbf {\bibinfo {volume} {10}},\ \bibinfo {pages} {104} (\bibinfo
  {year} {2018})},\ \Eprint {http://arxiv.org/abs/1807.06547} {arXiv:1807.06547
  [hep-ph]} \BibitemShut {NoStop}%
\bibitem [{\citenamefont {Mandal}(2018)}]{Mandal:2018czf}%
  \BibitemOpen
  \bibfield  {author} {\bibinfo {author} {\bibfnamefont {R.}~\bibnamefont
  {Mandal}},\ }\href {\doibase 10.1140/epjc/s10052-018-6192-3} {\bibfield
  {journal} {\bibinfo  {journal} {Eur. Phys. J. C}\ }\textbf {\bibinfo {volume}
  {78}},\ \bibinfo {pages} {726} (\bibinfo {year} {2018})},\ \Eprint
  {http://arxiv.org/abs/1808.07844} {arXiv:1808.07844 [hep-ph]} \BibitemShut
  {NoStop}%
\bibitem [{\citenamefont {Belanger}\ \emph {et~al.}(2022)\citenamefont
  {Belanger} \emph {et~al.}}]{Belanger:2021smw}%
  \BibitemOpen
  \bibfield  {author} {\bibinfo {author} {\bibfnamefont {G.}~\bibnamefont
  {Belanger}} \emph {et~al.},\ }\href {\doibase 10.1007/JHEP02(2022)042}
  {\bibfield  {journal} {\bibinfo  {journal} {JHEP}\ }\textbf {\bibinfo
  {volume} {02}},\ \bibinfo {pages} {042} (\bibinfo {year} {2022})},\ \Eprint
  {http://arxiv.org/abs/2111.08027} {arXiv:2111.08027 [hep-ph]} \BibitemShut
  {NoStop}%
\bibitem [{\citenamefont {Queiroz}\ \emph {et~al.}(2015)\citenamefont
  {Queiroz}, \citenamefont {Sinha},\ and\ \citenamefont
  {Strumia}}]{Queiroz:2014pra}%
  \BibitemOpen
  \bibfield  {author} {\bibinfo {author} {\bibfnamefont {F.~S.}\ \bibnamefont
  {Queiroz}}, \bibinfo {author} {\bibfnamefont {K.}~\bibnamefont {Sinha}}, \
  and\ \bibinfo {author} {\bibfnamefont {A.}~\bibnamefont {Strumia}},\ }\href
  {\doibase 10.1103/PhysRevD.91.035006} {\bibfield  {journal} {\bibinfo
  {journal} {Phys. Rev. D}\ }\textbf {\bibinfo {volume} {91}},\ \bibinfo
  {pages} {035006} (\bibinfo {year} {2015})},\ \Eprint
  {http://arxiv.org/abs/1409.6301} {arXiv:1409.6301 [hep-ph]} \BibitemShut
  {NoStop}%
\bibitem [{\citenamefont {D'Eramo}\ \emph {et~al.}(2021)\citenamefont
  {D'Eramo}, \citenamefont {Ko\v{s}nik}, \citenamefont {Pobbe}, \citenamefont
  {Smolkovi\v{c}},\ and\ \citenamefont {Sumensari}}]{DEramo:2020sqv}%
  \BibitemOpen
  \bibfield  {author} {\bibinfo {author} {\bibfnamefont {F.}~\bibnamefont
  {D'Eramo}}, \bibinfo {author} {\bibfnamefont {N.}~\bibnamefont {Ko\v{s}nik}},
  \bibinfo {author} {\bibfnamefont {F.}~\bibnamefont {Pobbe}}, \bibinfo
  {author} {\bibfnamefont {A.}~\bibnamefont {Smolkovi\v{c}}}, \ and\ \bibinfo
  {author} {\bibfnamefont {O.}~\bibnamefont {Sumensari}},\ }\href {\doibase
  10.1103/PhysRevD.104.015035} {\bibfield  {journal} {\bibinfo  {journal}
  {Phys. Rev. D}\ }\textbf {\bibinfo {volume} {104}},\ \bibinfo {pages}
  {015035} (\bibinfo {year} {2021})},\ \Eprint
  {http://arxiv.org/abs/2012.05743} {arXiv:2012.05743 [hep-ph]} \BibitemShut
  {NoStop}%
\bibitem [{\citenamefont {Belfatto}\ \emph {et~al.}(2021)\citenamefont
  {Belfatto}, \citenamefont {Buttazzo}, \citenamefont {Gross}, \citenamefont
  {Panci}, \citenamefont {Strumia}, \citenamefont {Vignaroli}, \citenamefont
  {Vittorio},\ and\ \citenamefont {Watanabe}}]{Belfatto:2021ats}%
  \BibitemOpen
  \bibfield  {author} {\bibinfo {author} {\bibfnamefont {B.}~\bibnamefont
  {Belfatto}}, \bibinfo {author} {\bibfnamefont {D.}~\bibnamefont {Buttazzo}},
  \bibinfo {author} {\bibfnamefont {C.}~\bibnamefont {Gross}}, \bibinfo
  {author} {\bibfnamefont {P.}~\bibnamefont {Panci}}, \bibinfo {author}
  {\bibfnamefont {A.}~\bibnamefont {Strumia}}, \bibinfo {author} {\bibfnamefont
  {N.}~\bibnamefont {Vignaroli}}, \bibinfo {author} {\bibfnamefont
  {L.}~\bibnamefont {Vittorio}}, \ and\ \bibinfo {author} {\bibfnamefont
  {R.}~\bibnamefont {Watanabe}},\ }\href@noop {} {\  (\bibinfo {year}
  {2021})},\ \Eprint {http://arxiv.org/abs/2111.14808} {arXiv:2111.14808
  [hep-ph]} \BibitemShut {NoStop}%
\bibitem [{\citenamefont {Baker}\ \emph {et~al.}(2021)\citenamefont {Baker},
  \citenamefont {Faroughy},\ and\ \citenamefont
  {Trifinopoulos}}]{Baker:2021llj}%
  \BibitemOpen
  \bibfield  {author} {\bibinfo {author} {\bibfnamefont {M.~J.}\ \bibnamefont
  {Baker}}, \bibinfo {author} {\bibfnamefont {D.~A.}\ \bibnamefont {Faroughy}},
  \ and\ \bibinfo {author} {\bibfnamefont {S.}~\bibnamefont {Trifinopoulos}},\
  }\href {\doibase 10.1007/JHEP11(2021)084} {\bibfield  {journal} {\bibinfo
  {journal} {JHEP}\ }\textbf {\bibinfo {volume} {11}},\ \bibinfo {pages} {084}
  (\bibinfo {year} {2021})},\ \Eprint {http://arxiv.org/abs/2109.08689}
  {arXiv:2109.08689 [hep-ph]} \BibitemShut {NoStop}%
\bibitem [{\citenamefont {Capdevila}\ \emph {et~al.}(2018)\citenamefont
  {Capdevila}, \citenamefont {Crivellin}, \citenamefont {Descotes-Genon},
  \citenamefont {Matias},\ and\ \citenamefont {Virto}}]{Capdevila:2017bsm}%
  \BibitemOpen
  \bibfield  {author} {\bibinfo {author} {\bibfnamefont {B.}~\bibnamefont
  {Capdevila}}, \bibinfo {author} {\bibfnamefont {A.}~\bibnamefont
  {Crivellin}}, \bibinfo {author} {\bibfnamefont {S.}~\bibnamefont
  {Descotes-Genon}}, \bibinfo {author} {\bibfnamefont {J.}~\bibnamefont
  {Matias}}, \ and\ \bibinfo {author} {\bibfnamefont {J.}~\bibnamefont
  {Virto}},\ }\href {\doibase 10.1007/JHEP01(2018)093} {\bibfield  {journal}
  {\bibinfo  {journal} {JHEP}\ }\textbf {\bibinfo {volume} {01}},\ \bibinfo
  {pages} {093} (\bibinfo {year} {2018})},\ \Eprint
  {http://arxiv.org/abs/1704.05340} {arXiv:1704.05340 [hep-ph]} \BibitemShut
  {NoStop}%
\bibitem [{\citenamefont {Alguer\'o}\ \emph {et~al.}(2019)\citenamefont
  {Alguer\'o}, \citenamefont {Capdevila}, \citenamefont {Crivellin},
  \citenamefont {Descotes-Genon}, \citenamefont {Masjuan}, \citenamefont
  {Matias}, \citenamefont {Novoa~Brunet},\ and\ \citenamefont
  {Virto}}]{Alguero:2019ptt}%
  \BibitemOpen
  \bibfield  {author} {\bibinfo {author} {\bibfnamefont {M.}~\bibnamefont
  {Alguer\'o}}, \bibinfo {author} {\bibfnamefont {B.}~\bibnamefont
  {Capdevila}}, \bibinfo {author} {\bibfnamefont {A.}~\bibnamefont
  {Crivellin}}, \bibinfo {author} {\bibfnamefont {S.}~\bibnamefont
  {Descotes-Genon}}, \bibinfo {author} {\bibfnamefont {P.}~\bibnamefont
  {Masjuan}}, \bibinfo {author} {\bibfnamefont {J.}~\bibnamefont {Matias}},
  \bibinfo {author} {\bibfnamefont {M.}~\bibnamefont {Novoa~Brunet}}, \ and\
  \bibinfo {author} {\bibfnamefont {J.}~\bibnamefont {Virto}},\ }\href
  {\doibase 10.1140/epjc/s10052-019-7216-3} {\bibfield  {journal} {\bibinfo
  {journal} {Eur. Phys. J. C}\ }\textbf {\bibinfo {volume} {79}},\ \bibinfo
  {pages} {714} (\bibinfo {year} {2019})},\ \bibinfo {note} {[Addendum:
  Eur.Phys.J.C 80, 511 (2020)]},\ \Eprint {http://arxiv.org/abs/1903.09578}
  {arXiv:1903.09578 [hep-ph]} \BibitemShut {NoStop}%
\bibitem [{\citenamefont {Ciuchini}\ \emph
  {et~al.}(2021{\natexlab{b}})\citenamefont {Ciuchini}, \citenamefont {Fedele},
  \citenamefont {Franco}, \citenamefont {Paul}, \citenamefont {Silvestrini},\
  and\ \citenamefont {Valli}}]{Ciuchini:2020gvn}%
  \BibitemOpen
  \bibfield  {author} {\bibinfo {author} {\bibfnamefont {M.}~\bibnamefont
  {Ciuchini}}, \bibinfo {author} {\bibfnamefont {M.}~\bibnamefont {Fedele}},
  \bibinfo {author} {\bibfnamefont {E.}~\bibnamefont {Franco}}, \bibinfo
  {author} {\bibfnamefont {A.}~\bibnamefont {Paul}}, \bibinfo {author}
  {\bibfnamefont {L.}~\bibnamefont {Silvestrini}}, \ and\ \bibinfo {author}
  {\bibfnamefont {M.}~\bibnamefont {Valli}},\ }\href {\doibase
  10.1103/PhysRevD.103.015030} {\bibfield  {journal} {\bibinfo  {journal}
  {Phys. Rev. D}\ }\textbf {\bibinfo {volume} {103}},\ \bibinfo {pages}
  {015030} (\bibinfo {year} {2021}{\natexlab{b}})},\ \Eprint
  {http://arxiv.org/abs/2011.01212} {arXiv:2011.01212 [hep-ph]} \BibitemShut
  {NoStop}%
\bibitem [{\citenamefont {Zyla}\ \emph {et~al.}(2020)\citenamefont {Zyla} \emph
  {et~al.}}]{Zyla:2020zbs}%
  \BibitemOpen
  \bibfield  {author} {\bibinfo {author} {\bibfnamefont {P.~A.}\ \bibnamefont
  {Zyla}} \emph {et~al.} (\bibinfo {collaboration} {Particle Data Group}),\
  }\href {\doibase 10.1093/ptep/ptaa104} {\bibfield  {journal} {\bibinfo
  {journal} {PTEP}\ }\textbf {\bibinfo {volume} {2020}},\ \bibinfo {pages}
  {083C01} (\bibinfo {year} {2020})}\BibitemShut {NoStop}%
\bibitem [{\citenamefont {Grygier}\ \emph {et~al.}(2017)\citenamefont {Grygier}
  \emph {et~al.}}]{Belle:2017oht}%
  \BibitemOpen
  \bibfield  {author} {\bibinfo {author} {\bibfnamefont {J.}~\bibnamefont
  {Grygier}} \emph {et~al.} (\bibinfo {collaboration} {Belle}),\ }\href
  {\doibase 10.1103/PhysRevD.96.091101} {\bibfield  {journal} {\bibinfo
  {journal} {Phys. Rev. D}\ }\textbf {\bibinfo {volume} {96}},\ \bibinfo
  {pages} {091101} (\bibinfo {year} {2017})},\ \bibinfo {note} {[Addendum:
  Phys.Rev.D 97, 099902 (2018)]},\ \Eprint {http://arxiv.org/abs/1702.03224}
  {arXiv:1702.03224 [hep-ex]} \BibitemShut {NoStop}%
\bibitem [{\citenamefont {Abe}\ \emph {et~al.}(2010)\citenamefont {Abe} \emph
  {et~al.}}]{Belle-II:2010dht}%
  \BibitemOpen
  \bibfield  {author} {\bibinfo {author} {\bibfnamefont {T.}~\bibnamefont
  {Abe}} \emph {et~al.} (\bibinfo {collaboration} {Belle-II}),\ }\href@noop {}
  {\  (\bibinfo {year} {2010})},\ \Eprint {http://arxiv.org/abs/1011.0352}
  {arXiv:1011.0352 [physics.ins-det]} \BibitemShut {NoStop}%
\bibitem [{\citenamefont {Angelescu}\ \emph {et~al.}(2018)\citenamefont
  {Angelescu}, \citenamefont {Be\v{c}irevi\'c}, \citenamefont {Faroughy},\ and\
  \citenamefont {Sumensari}}]{Angelescu:2018tyl}%
  \BibitemOpen
  \bibfield  {author} {\bibinfo {author} {\bibfnamefont {A.}~\bibnamefont
  {Angelescu}}, \bibinfo {author} {\bibfnamefont {D.}~\bibnamefont
  {Be\v{c}irevi\'c}}, \bibinfo {author} {\bibfnamefont {D.~A.}\ \bibnamefont
  {Faroughy}}, \ and\ \bibinfo {author} {\bibfnamefont {O.}~\bibnamefont
  {Sumensari}},\ }\href {\doibase 10.1007/JHEP10(2018)183} {\bibfield
  {journal} {\bibinfo  {journal} {JHEP}\ }\textbf {\bibinfo {volume} {10}},\
  \bibinfo {pages} {183} (\bibinfo {year} {2018})},\ \Eprint
  {http://arxiv.org/abs/1808.08179} {arXiv:1808.08179 [hep-ph]} \BibitemShut
  {NoStop}%
\bibitem [{CMS(2019)}]{CMS-PAS-LUM-18-002}%
  \BibitemOpen
  \href {https://cds.cern.ch/record/2676164} {\emph {\bibinfo {title} {{CMS
  luminosity measurement for the 2018 data-taking period at $\sqrt{s} =
  13~\mathrm{TeV}$}}}},\ \bibinfo {type} {Tech. Rep.}\ (\bibinfo  {institution}
  {CERN},\ \bibinfo {address} {Geneva},\ \bibinfo {year} {2019})\BibitemShut
  {NoStop}%
\bibitem [{\citenamefont {Profumo}(2020)}]{Profumo:2020zgi}%
  \BibitemOpen
  \bibfield  {author} {\bibinfo {author} {\bibfnamefont {S.}~\bibnamefont
  {Profumo}},\ }\href {\doibase 10.1103/PhysRevD.102.035008} {\bibfield
  {journal} {\bibinfo  {journal} {Phys. Rev. D}\ }\textbf {\bibinfo {volume}
  {102}},\ \bibinfo {pages} {035008} (\bibinfo {year} {2020})},\ \Eprint
  {http://arxiv.org/abs/2005.02512} {arXiv:2005.02512 [hep-th]} \BibitemShut
  {NoStop}%
\bibitem [{\citenamefont {Cirelli}\ \emph {et~al.}(2006)\citenamefont
  {Cirelli}, \citenamefont {Fornengo},\ and\ \citenamefont
  {Strumia}}]{Cirelli:2005uq}%
  \BibitemOpen
  \bibfield  {author} {\bibinfo {author} {\bibfnamefont {M.}~\bibnamefont
  {Cirelli}}, \bibinfo {author} {\bibfnamefont {N.}~\bibnamefont {Fornengo}}, \
  and\ \bibinfo {author} {\bibfnamefont {A.}~\bibnamefont {Strumia}},\ }\href
  {\doibase 10.1016/j.nuclphysb.2006.07.012} {\bibfield  {journal} {\bibinfo
  {journal} {Nucl. Phys. B}\ }\textbf {\bibinfo {volume} {753}},\ \bibinfo
  {pages} {178} (\bibinfo {year} {2006})},\ \Eprint
  {http://arxiv.org/abs/hep-ph/0512090} {arXiv:hep-ph/0512090} \BibitemShut
  {NoStop}%
\bibitem [{\citenamefont {Barducci}\ \emph {et~al.}(2018)\citenamefont
  {Barducci}, \citenamefont {Belanger}, \citenamefont {Bernon}, \citenamefont
  {Boudjema}, \citenamefont {Da~Silva}, \citenamefont {Kraml}, \citenamefont
  {Laa},\ and\ \citenamefont {Pukhov}}]{Barducci:2016pcb}%
  \BibitemOpen
  \bibfield  {author} {\bibinfo {author} {\bibfnamefont {D.}~\bibnamefont
  {Barducci}}, \bibinfo {author} {\bibfnamefont {G.}~\bibnamefont {Belanger}},
  \bibinfo {author} {\bibfnamefont {J.}~\bibnamefont {Bernon}}, \bibinfo
  {author} {\bibfnamefont {F.}~\bibnamefont {Boudjema}}, \bibinfo {author}
  {\bibfnamefont {J.}~\bibnamefont {Da~Silva}}, \bibinfo {author}
  {\bibfnamefont {S.}~\bibnamefont {Kraml}}, \bibinfo {author} {\bibfnamefont
  {U.}~\bibnamefont {Laa}}, \ and\ \bibinfo {author} {\bibfnamefont
  {A.}~\bibnamefont {Pukhov}},\ }\href {\doibase 10.1016/j.cpc.2017.08.028}
  {\bibfield  {journal} {\bibinfo  {journal} {Comput. Phys. Commun.}\ }\textbf
  {\bibinfo {volume} {222}},\ \bibinfo {pages} {327} (\bibinfo {year}
  {2018})},\ \Eprint {http://arxiv.org/abs/1606.03834} {arXiv:1606.03834
  [hep-ph]} \BibitemShut {NoStop}%
\bibitem [{\citenamefont {Aghanim}\ \emph {et~al.}(2020)\citenamefont {Aghanim}
  \emph {et~al.}}]{Planck:2018vyg}%
  \BibitemOpen
  \bibfield  {author} {\bibinfo {author} {\bibfnamefont {N.}~\bibnamefont
  {Aghanim}} \emph {et~al.} (\bibinfo {collaboration} {Planck}),\ }\href
  {\doibase 10.1051/0004-6361/201833910} {\bibfield  {journal} {\bibinfo
  {journal} {Astron. Astrophys.}\ }\textbf {\bibinfo {volume} {641}},\ \bibinfo
  {pages} {A6} (\bibinfo {year} {2020})},\ \bibinfo {note} {[Erratum:
  Astron.Astrophys. 652, C4 (2021)]},\ \Eprint
  {http://arxiv.org/abs/1807.06209} {arXiv:1807.06209 [astro-ph.CO]}
  \BibitemShut {NoStop}%
\bibitem [{\citenamefont {Strigari}(2009)}]{Strigari:2009bq}%
  \BibitemOpen
  \bibfield  {author} {\bibinfo {author} {\bibfnamefont {L.~E.}\ \bibnamefont
  {Strigari}},\ }\href {\doibase 10.1088/1367-2630/11/10/105011} {\bibfield
  {journal} {\bibinfo  {journal} {New J. Phys.}\ }\textbf {\bibinfo {volume}
  {11}},\ \bibinfo {pages} {105011} (\bibinfo {year} {2009})},\ \Eprint
  {http://arxiv.org/abs/0903.3630} {arXiv:0903.3630 [astro-ph.CO]} \BibitemShut
  {NoStop}%
\bibitem [{\citenamefont {Ackermann}\ \emph {et~al.}(2015)\citenamefont
  {Ackermann} \emph {et~al.}}]{Fermi-LAT:2015att}%
  \BibitemOpen
  \bibfield  {author} {\bibinfo {author} {\bibfnamefont {M.}~\bibnamefont
  {Ackermann}} \emph {et~al.} (\bibinfo {collaboration} {Fermi-LAT}),\ }\href
  {\doibase 10.1103/PhysRevLett.115.231301} {\bibfield  {journal} {\bibinfo
  {journal} {Phys. Rev. Lett.}\ }\textbf {\bibinfo {volume} {115}},\ \bibinfo
  {pages} {231301} (\bibinfo {year} {2015})},\ \Eprint
  {http://arxiv.org/abs/1503.02641} {arXiv:1503.02641 [astro-ph.HE]}
  \BibitemShut {NoStop}%
\bibitem [{\citenamefont {Abdallah}\ \emph {et~al.}(2016)\citenamefont
  {Abdallah} \emph {et~al.}}]{HESS:2016mib}%
  \BibitemOpen
  \bibfield  {author} {\bibinfo {author} {\bibfnamefont {H.}~\bibnamefont
  {Abdallah}} \emph {et~al.} (\bibinfo {collaboration} {H.E.S.S.}),\ }\href
  {\doibase 10.1103/PhysRevLett.117.111301} {\bibfield  {journal} {\bibinfo
  {journal} {Phys. Rev. Lett.}\ }\textbf {\bibinfo {volume} {117}},\ \bibinfo
  {pages} {111301} (\bibinfo {year} {2016})},\ \Eprint
  {http://arxiv.org/abs/1607.08142} {arXiv:1607.08142 [astro-ph.HE]}
  \BibitemShut {NoStop}%
\bibitem [{\citenamefont {Drlica-Wagner}\ \emph {et~al.}(2019)\citenamefont
  {Drlica-Wagner} \emph {et~al.}}]{LSSTDarkMatterGroup:2019mwo}%
  \BibitemOpen
  \bibfield  {author} {\bibinfo {author} {\bibfnamefont {A.}~\bibnamefont
  {Drlica-Wagner}} \emph {et~al.} (\bibinfo {collaboration} {LSST Dark Matter
  Group}),\ }\href@noop {} {\  (\bibinfo {year} {2019})},\ \Eprint
  {http://arxiv.org/abs/1902.01055} {arXiv:1902.01055 [astro-ph.CO]}
  \BibitemShut {NoStop}%
\bibitem [{\citenamefont {Acharyya}\ \emph {et~al.}(2021)\citenamefont
  {Acharyya} \emph {et~al.}}]{CTA:2020qlo}%
  \BibitemOpen
  \bibfield  {author} {\bibinfo {author} {\bibfnamefont {A.}~\bibnamefont
  {Acharyya}} \emph {et~al.} (\bibinfo {collaboration} {CTA}),\ }\href
  {\doibase 10.1088/1475-7516/2021/01/057} {\bibfield  {journal} {\bibinfo
  {journal} {JCAP}\ }\textbf {\bibinfo {volume} {01}},\ \bibinfo {pages} {057}
  (\bibinfo {year} {2021})},\ \Eprint {http://arxiv.org/abs/2007.16129}
  {arXiv:2007.16129 [astro-ph.HE]} \BibitemShut {NoStop}%
\bibitem [{\citenamefont {Abdalla}\ \emph {et~al.}(2021)\citenamefont {Abdalla}
  \emph {et~al.}}]{HESS:2021pgk}%
  \BibitemOpen
  \bibfield  {author} {\bibinfo {author} {\bibfnamefont {H.}~\bibnamefont
  {Abdalla}} \emph {et~al.} (\bibinfo {collaboration} {H.E.S.S.}),\ }\href
  {\doibase 10.3847/1538-4357/abff59} {\bibfield  {journal} {\bibinfo
  {journal} {Astrophys. J.}\ }\textbf {\bibinfo {volume} {918}},\ \bibinfo
  {pages} {17} (\bibinfo {year} {2021})},\ \Eprint
  {http://arxiv.org/abs/2106.00551} {arXiv:2106.00551 [astro-ph.HE]}
  \BibitemShut {NoStop}%
\end{thebibliography}%

\end{document}